\documentclass[twocolumn,]{aastex701}

\usepackage{amsmath}
\usepackage{booktabs}
\usepackage{multirow}
\usepackage{threeparttable}
\usepackage{makecell}
\usepackage[table]{xcolor}
\usepackage{soul}

\begin{document}




\defcitealias{Ling2024}{L24}

\title{Mid-IR luminosity functions: inferred dusty cosmic star-formation and black hole accretion histories from the JWST SMILES}

\author[0000-0002-3738-1834]{Chih-Teng Ling}
\affiliation{Institute of Astronomy, National Tsing Hua University, 101, Section 2. Kuang-Fu Road, Hsinchu, 30013, Taiwan (R.O.C.)}
\email[show]{s106022129@m106.nthu.edu.tw}

\author[0000-0002-6821-8669]{Tomotsugu Goto}
\affiliation{Institute of Astronomy, National Tsing Hua University, 101, Section 2. Kuang-Fu Road, Hsinchu, 30013, Taiwan (R.O.C.)}
\affiliation{Department of Physics, National Tsing Hua University, 101, Section 2. Kuang-Fu Road, Hsinchu, 30013, Taiwan (R.O.C.)}
\email[]{}

\author[0000-0001-9970-8145]{Seong Jin Kim}
\affiliation{Institute of Astronomy, National Tsing Hua University, 101, Section 2. Kuang-Fu Road, Hsinchu, 30013, Taiwan (R.O.C.)}
\email[]{}

\author[0000-0003-4258-2996]{Cossas K.-W. Wu}
\affiliation{Institute of Astronomy, National Tsing Hua University, 101, Section 2. Kuang-Fu Road, Hsinchu, 30013, Taiwan (R.O.C.)}
\email[]{}

\author[0009-0004-9353-7065]{Amos Y.-A. Chen}
\affiliation{Department of Physics, National Tsing Hua University, 101, Section 2. Kuang-Fu Road, Hsinchu, 30013, Taiwan (R.O.C.)}
\email[]{}

\author[0000-0002-9119-2313]{Ece Kilerci}
\affiliation{Department of Astronomy and Space Sciences, Science Faculty, \.{I}stanbul University, Beyaz{\i}t 34119, \.Istanbul, T\"{u}rkiye}
\email[]{}

\author[0000-0001-7228-1428]{Tetsuya Hashimoto}
\affiliation{Department of Physics, National Chung Hsing University, 145, Xingda Road, Taichung, 40227, Taiwan (R.O.C.)}
\email[]{}

\author[0009-0009-2940-0861]{Terry Long Phan}
\affiliation{Institute of Astronomy, National Tsing Hua University, 101, Section 2. Kuang-Fu Road, Hsinchu, 30013, Taiwan (R.O.C.)}
\email[]{}

\correspondingauthor{Chih-Teng Ling}
\begin{abstract}
Mid-infrared (mid-IR) observations are crucial for understanding galaxy evolution, tracing star formation, and active galactic nuclei (AGN) activity via dust emission. 
This work presents mid-IR galaxy luminosity functions (LFs) at $0.5 < z < 6$, derived from the JWST Systematic Mid-infrared Instrument Legacy Extragalactic Survey (SMILES) program.
We combine 8 MIRI bands ($5-25$ $\mu$m) of SMILES and archival 23-band HST+JWST NIRCam photometry to construct an extensive catalog containing 2,813 galaxies with sub-$\mu$Jy level completeness in the mid-IR.
We obtain monochromatic (in 5.6, 7.7, 10, 12.8, 15, 18, 21, and 25.5 $\mu$m), $L_{\rm IR}$, and AGN LFs, with a limiting luminosity down to $10^{9.5}$ $L_\odot$ at $z=0.5-1.0$, $\sim10^{10.5}$ $L_\odot$ at $z=2.0-4.0$, and to $\sim10^{11}$ $L_\odot$ at $z=4.0-6.0$.
With the unprecedented sensitivity and resolution of JWST, here we better constrain the faint-end slope and its evolution of the mid-IR LFs, quantifying the dusty cosmic star formation and black hole accretion histories out to $z \sim 5$. 
These results provide essential insights to refine our understanding of the obscured star formation and galaxy-AGN co-evolution over cosmic time. 
\end{abstract}



\section{Introduction}
Understanding the evolution of galaxies and the cosmic processes that drive their growth, such as star formation (SF) and active galactic nuclei (AGN) activity, is a fundamental goal in modern astrophysics. 
A complete picture requires tracing not only the unobscured stellar light, predominantly in the optical and ultraviolet (UV), but also the vast energy output reprocessed by interstellar dust.
This dust emission, which peaks in the infrared (IR), is powered by two fundamental processes: intense star formation within molecular clouds and accretion onto supermassive black holes (SMBHs) in AGN. 
At the peak of cosmic activity, commonly noted as "cosmic noon" ($z \sim 1-3$), a significant fraction of star formation and black hole growth was obscured by dust \citep[e.g.,][]{Madau2014ARA&A..52..415M, Hickox2018ARA&A..56..625H, Vijarnwannaluk2022ApJ...941...97V}. 
Even towards higher redshift $z \sim 7$, the dust-obscured SF can contribute as high as ~30\% of total SF \citep[e.g.,][]{Zavala2021ApJ...909..165Z, Algera2023MNRAS.518.6142A}.
These phenomena suggest that characterizing the IR emission properties of galaxies is essential for a comprehensive census of these evolutionary drivers.

The luminosity function (LF), which describes the number density of galaxies as a function of their luminosity, is a powerful statistical tool for quantifying this evolution.
IR space telescopes such as the AKARI, Spitzer, and Herschel have been refreshing our understanding of the dusty universe for the past decades. 
Benefiting from the large-scale surveys, they are able to construct IR LFs out to $z>1$ epoch for the first time \citep[e.g.,][]{LeFloch2005ApJ...632..169L, Caputi2007, Goto2010A&A...514A...6G, Gruppioni2013MNRAS.432...23G, Magnelli2013A&A...553A.132M}, revealing a strong cosmic evolution where the number density of luminous and ultra-luminous infrared galaxies (LIRGs and ULIRGs) increases dramatically from the local universe to cosmic noon, confirming the significance of these IR populations in the grand picture of cosmic galaxy evolution. 

Mid-infrared (mid-IR) observations are particularly critical in these efforts due to their capability to effectively track star formation and AGN-heated dust emission features, which are missed by optical or UV surveys \citep[e.g.][]{Goto2011MNRAS.410..573G, Goto2011MNRAS.414.1903G, Goto2015MNRAS.452.1684G, Kim2015MNRAS.454.1573K}.
However, because of limitations in sensitivity and angular resolution, these works can only depict the brightest sources generally, leaving the faint end of the LF ($L_{\rm IR} \lesssim 10^{11} L_\odot$) poorly constrained, especially at $z>1$. 
It has been common to assume a fixed value of the faint-end slope to extrapolate LF and estimate the infrared luminosity density (known as the cosmic star formation rate density, CSFRD).
Therefore, our understanding of the dominant mode of star formation and its evolution could easily be biased and limited to brighter galaxies.
Similarly, while deep X-ray surveys have provided an important pathway for studying AGN evolution \citep[e.g.,][]{Ueda2003ApJ...598..886U, Barger2005AJ....129..578B, Aird2015MNRAS.451.1892A}, a complete census requires probing the significant population of heavily obscured AGN that are faint or invisible in X-rays but only bright in MIR \citep{Lacy2015ApJ...802..102L}.

The advent of the James Webb Space Telescope (JWST), with its unprecedented sensitivity and spatial resolution, the Mid-Infrared Instrument \citep[MIRI;][]{Rieke2015PASP..127..584R} is capable of probing the faint, distant universe with remarkable clarity, orders of magnitude deeper than its predecessors. 
The first wave of MIRI results mainly from the JWST Cycle 1 program, Cosmic Evolution Early Release Science Survey \citep[CEERS;] []{CEERS_2017jwst.prop.1345F}: dusty, low-luminosity AGNs that were missed by AKARI and Spitzer now comprise a new population in the high-$z$ universe \citep[e.g.,][]{Yang2023ApJ...950L...5Y, Kirkpatrick2023ApJ...959L...7K, Barro2024ApJ...963..128B, Lin2024, Chien2024MNRAS.532..719C}; the deepest IR source count \citep{Wu2023MNRAS.523.5187W, Yang2023ApJ...956L..12Y}, the first JWST-based IR LF (\citealt{Ling2024}, hereafter \citetalias{Ling2024}) and MIR-selected AGN LF \citep{Hsieh2025} 
are provided, pushing our knowledge of galaxy evolution and black hole history to the fainter and further reaches than ever before.

Despite this rapid progress, these initial studies have been limited by the relatively small areas of the first deep fields (few arcmin$^2$), making their results susceptible to small-number statistics.
The JWST Cycle 3 Systematic Mid-infrared Instrument Legacy Extragalactic Survey \citep[SMILES;][]{SMILES_Rieke2024, SMILES_Alberts2024} is the most complete MIRI program to date, utilizing all 8 of MIRI broadband filters with a wider survey area of 34.5 arcmin$^2$, four times that of CEERS.
Preliminary source count results from SMILES \citep{Stone2024ApJ...972...62S, Sajkov2024ApJ...977..115S} have demonstrated the remarkable depth and sensitivity of the new JWST data and confirmed the consistency between observations.
However, a detailed analysis to convert these source counts into solid luminosity functions has not yet been undertaken.

The primary goals of this work are: 1) to derive robust monochromatic and infrared ($L_{\rm IR}$) LF for JWST galaxies with the larger sample size, improving earlier JWST efforts \citepalias{Ling2024}, 2) to precisely constrain the faint-end slope of the LFs and investigate its evolution with cosmic time for the first time, and 3) to trace the evolution of the obscured CSFRD and black hole accretion density based on newly detected faint galaxies, providing crucial new constraints on our models of galaxy evolution.

This paper is structured as follows. 
In Section \ref{sec:data}, we describe the SMILES and JADES datasets and the construction of our multi-wavelength catalog. 
Section \ref{sec:analysis} details our methodology for photometric redshift estimation with EAZY and spectral energy distribution (SED) fitting with CIGALE. 
In Section \ref{sec:LF}, we present and discuss our derived monochromatic, $L_{\rm IR}$, and AGN luminosity functions, as well as their functional fits. 
In Section \ref{sec:ld}, we present the derived cosmic star formation density and black hole accretion rate history. 
Our main findings are summarized in Section \ref{sec:summary}. 
Throughout the paper, we adopt the Planck18 cosmology \citep{Planck2020A&A...641A...6P} with ($\Omega_{m}$, $\Omega_{\Lambda}$, $\Omega_{b}$, $h)=(0.310, 0.689, 0.0490, 0.677)$.

\section{Data}
\label{sec:data}
\subsection{SMILES}
\label{sec:smiles}
The Systematic Mid-infrared Instrument Legacy Extragalactic Survey \citep[SMILES, PID 1207;][]{SMILES_Rieke2024} is the largest MIRI imaging survey to date that utilizes 8 MIRI bands (F560W, F770W, F1000W, F1280W, F1500W, F1800W, F2100W, F2550W; $5.6 - 25.5 \mu$m).
The survey covers a total area of $\sim34.5$ arcmin$^2$ in the GOODS-S / HUDF field, with an exposure time of approximately 650 - 2100 seconds per band. 
Despite the relatively short exposure time, SMILES is the only JWST survey that provides eight contiguous MIRI filters in mid-IR, making it the most comprehensive dataset for studying the spectral energy distribution (SED) of mid-IR galaxies.

This work uses the photometric catalog from the SMILES initial data release \citep{SMILES_Alberts2024}. 
By aligning the SMILES images to the JADES NIRCam catalog \citep{JADES_Rieke2023ApJS..269...16R}, the final astrometric accuracy is $0.1-0.2$ ($0.4$) arcsec for F560W-F2100W (F2550W).
SMILES photometry is performed with a modified version of the JADES photometric pipeline \citep{JADES_Rieke2023ApJS..269...16R}, explicitly tailored to MIRI data.
\cite{SMILES_Alberts2024} adopt the F560W+F770W stack detection image to identify sources and measure their aperture photometry (including fixed circular apertures, and Kron aperture taken $2.5 \times$ scaled Kron radius) in all eight MIRI bands through a careful quality control process.
The photometry is then corrected for the aperture with custom PSFs generated from WebbPSF \citep{Perrin2014SPIE.9143E..3XP} and empirical PSFs derived from JWST commissioning images.
Their final catalog contains 3096 sources with SNR $>4$ in the F560W or F770W bands.

We find SMILES sources can reach 5$\sigma$ depth as well as 80\% completeness at sub-$\mu$Jy level \citep{Stone2024ApJ...972...62S} comparable to CEERS \citep[PID 1345;][]{CEERS_2017jwst.prop.1345F, Yang2023ApJ...956L..12Y}. 
A comparison between depth in SMILES and CEERS (taking MIRI1 pointing) data has been presented in Table \ref{tab:smiles}. 
As shown, the 5$\sigma$ limit achieved by SMILES is similar to or slightly better than CEERS, despite a generally shorter exposure time for SMILES and the shallower 5$\sigma$ value predicted by the Exposure Time Calculator (ETC).
A likely explanation for the improvement in the SMILES photometry \citep[$\sim2\times$ deeper sensitivity limit than ETC predictions,][]{SMILES_Alberts2024} could be differences in reduction pipelines, rather than intrinsic sky background or observation strategies.

Using ETC version 5.0, we calculated that the CEERS background emission is only $5-10\%$ stronger than SMILES in shorter MIRI bands, likely attributed to zodiacal light, but is insufficient to explain the discrepancy in observed depth. 
More significant differences could arise from the reduction processes. 
While both surveys follow similar image calibration steps, including warm pixel removal and super-background subtraction \citep[e.g.,][]{Yang2023ApJ...956L..12Y, Perez-Gonzalez2024ApJ...968....4P}, SMILES data reduction is performed using the JWST Calibration Pipeline v1.12.5 \citep{Bushouse2023zndo...7577320B}, a newer version compared to CEERS (v1.10.2). 
Improvements in the newer pipeline regarding detector artifact mitigation (such as cosmic ray showers) and 1/f noise correction are more likely to contribute to the enhanced quality of the final mosaics.

\subsection{JADES and ancillary data}
\label{sec:jades}
A wealth of extensive multi-wavelength data in the GOODS-S field complements SMILES.
In this work, we utilize the catalog from the JWST Advanced Deep Extragalactic Survey \citep[JADES, PID 1180;][]{JADES_2023arXiv230602465E} Data Release 3 \citep{JADES_Rieke2023ApJS..269...16R, JADES_2024A&A...690A.288B, JADES_DR3_2024arXiv240406531D}, the deepest ever near-IR imaging survey in the field.
JADES provides deep photometries for 9 NIRCam imaging bands (F090W - F444W) with a typical 5$\sigma$ depth of 29 AB magnitude.
Their catalog also includes 5 additional medium-band filters (F182M, F210M, F430M, F460M, F480M) from the JWST Extragalactic Medium-band Survey \citep[JEMS, PID 1963;][]{JEMS_2023ApJS..268...64W}. 
In addition to the NIRCam bands, JADES combines 9 bands of HST WFC3 / IR and ACS photometry based on imaging data from the CANDELS survey \citep{CANDELS_2011ApJS..197...35G, CANDELS_2011ApJS..197...36K}.
23 bands in total are available for over 45,000 sources in the JADES catalog.
We also match available spectroscopic redshifts (spec-$z$) from NIRSpec medium-resolution gratings, which form part of the JADES survey \citep{JADES_2024A&A...690A.288B, JADES_DR3_2024arXiv240406531D}.
Spec-$z$ information would be used later to validate our independent photometric redshift (photo-$z$) estimates.

\subsection{Catalog construction}
\label{sec:catalog}
\begin{figure*}
    \centering
    \includegraphics[width=\textwidth] {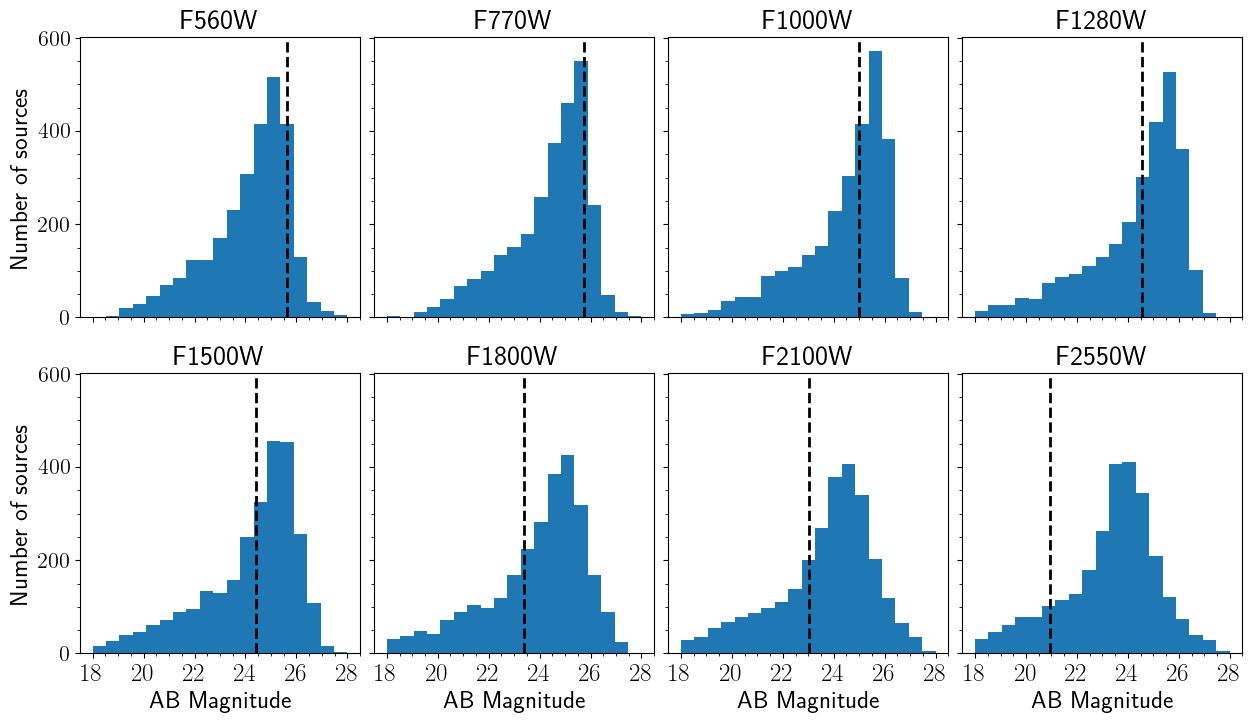}
    \caption{Magnitude histograms of our SMILES sample. 
    Each panel displays the distribution for one of the eight MIRI bands.
    The black dashed line represents the 80\% completeness limit of each band, converted to AB magnitude.}
    \label{fig:mag_hist}
\end{figure*}

We cross-match the SMILES and JADES photometric catalogs to create a comprehensive multi-wavelength catalog spanning from optical to mid-IR (observed frame, $0.4 - 25.5$ $\mu$m).
A matching radius of 0.2 arcsec is chosen as we find it to be the optimal value that ensures a sufficient number (compared to smaller matching radii of 0.1 or 0.15 arcsec) of robust source associations while being smaller than the smallest aperture (0.25 arcsec) used in the SMILES photometry, minimizing spurious matches due to different resolutions of instruments. 
Our final catalog with both SMILES and JADES photometry contains 2813 sources (96\% found within 0.1 arcsec). 
Of these, 276 ($\sim 10 \%$) sources have spec-$z$ from the NIRSpec grating line fluxes.
We visually inspect the 283 mismatched sources and find 64\% (180) of them to be from non-overlapping coverage, 20\% (57) of sources lacking NIRCam counterparts are those affected by nearby ($\sim1$ arcsec) bright NIR sources, resulting in missing NIRCam photometry, and 16\% (46) unreliable sources with low signal-to-noise ratio ($<5$ in either F560W or F770W kron photometry).
To compensate for the 57 sources without NIR photometry (1.8\% of the total), we multiplied the LF by a factor of 1/0.982. See Section \ref{sec:qc} for detailed justification.
The magnitude distribution of the catalog can be found in Figure~\ref{fig:mag_hist}.

\begin{deluxetable}{ccccc}
\tablewidth{0pt}
\tablecaption{Comparison of Band Depth in the SMILES and CEERS Observation}
\label{tab:smiles}
\tablehead{
\colhead{\textbf{Band}} & 
\multicolumn{2}{c}{\textbf{Exposure Time (s)}} & 
\multicolumn{2}{c}{\textbf{Measured (ETC Predicted)}} \\
\colhead{} & \colhead{} & \colhead{} & 
\multicolumn{2}{c}{\textbf{5$\sigma$ Limit (AB)}} \\ 
\colhead{} & 
\colhead{\textbf{SMILES}} & 
\colhead{\textbf{CEERS}} & 
\colhead{\textbf{SMILES}} & \colhead{\textbf{CEERS}}
}
\startdata
F770W & 866 & 1648 & 25.7 (24.6) & 25.6 (25.2)\\
F1000W &  644 & 1673 & 24.9 (24.1) & 24.8 (24.7)\\
F1280W &  755 & 1673 & 24.4 (23.6) & 24.2 (24.1)\\
F1500W &  1121 & 1673 & 24.2 (23.5) & 23.6 (23.8)\\
F1800W &  755 & 1698 & 23.3 (22.6) & 22.9 (22.9)\\
F2100W &  2187 & 4812 & 22.8 (22.6) & 22.2 (23.1)
\enddata
\end{deluxetable}

\section{Analysis}
\label{sec:analysis}
\subsection{Photometric Redshift}
The first step in deriving the luminosity function is to determine the redshift and fit the SED of the galaxies.  
Unlike \citetalias{Ling2024}, we divide this process into two steps to efficiently handle our larger catalog: first, we compute the photo-$z$ with EAZY \citep{Brammer2008ApJ...686.1503B}. 
Then, we use the output redshift as a fixed prior to perform the SED fitting with Code Investigating GALaxy Emission \citep[CIGALE,][]{Boquien2019A&A...622A.103B}.
This approach significantly reduces the time that CIGALE spends calculating redshift templates, allowing us to complete the fitting smoothly without sacrificing CIGALE's modeling capabilities.

EAZY performs a linear combination of user-supplied galaxy SED templates to fit the observed photometry, determining the best-fit redshift and its probability distribution function (PDF) for each object. 
We use the SED templates provided by \cite{Hainline2024ApJ...964...71H}, which include 9 built-in templates from EAZY v1.3 \citep{Brammer2008ApJ...686.1503B, Erb2010ApJ...719.1168E}, as well as 7 templates designed by \cite{Hainline2024ApJ...964...71H} for JADES galaxies.
These templates are based on the JAGUAR simulations \citep{Williams2018ApJS..236...33W} and are able to fit a broader range of galaxy colors, from dusty to UV-bright galaxies.

\begin{figure*}
    \centering
    \includegraphics[width=.45\textwidth] {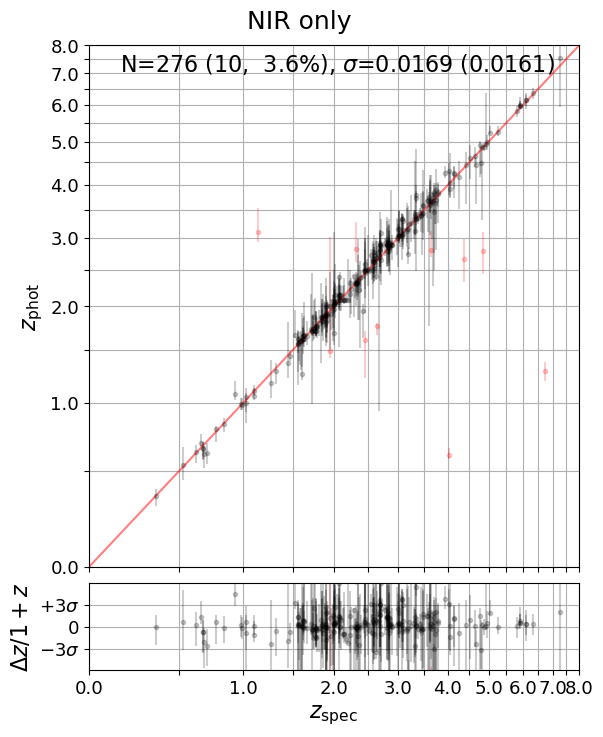}
    \includegraphics[width=.45\textwidth] {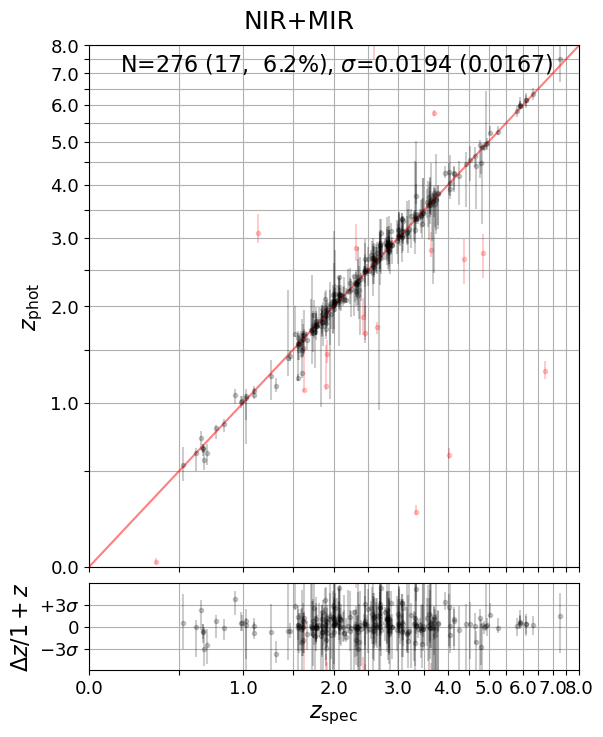}
    \caption{Comparison between spec-$z$ ($z_\mathrm{spec}$) and EAZY photo-$z$ ($z_\mathrm{phot}$) for 276 sources with available NIRSpec data.
    The left panel shows the results obtained using only NIRCam photometry (NIR only), while the right panel includes both NIRCam and MIRI photometry (NIR+MIR).
    Red data points indicate catastrophic outliers.
    The lower subpanels show the distribution of $\Delta z / (1+z)$, where $\Delta z = z_\mathrm{phot} - z_\mathrm{spec}$.
    The number of sources (N), the number of outliers and their fraction ($\eta$), and $\sigma_\mathrm{NMAD}$ (with and without outliers) are shown at the top.
    }
    \label{fig:eazy}
\end{figure*}

Our EAZY configuration follows \cite{Hainline2024ApJ...964...71H}, and \cite{JADES_Rieke2023ApJS..269...16R}, adopting Kron aperture photometry to accommodate galaxies of various sizes. 
We allow a redshift range of $0-10$ and add a systematic error of 5\%.
The results are evaluated by comparing the spec-$z$ of 276 sources (refer to Section \ref{sec:catalog}) with their photo-$z$. 
Figure \ref{fig:eazy} shows how including or excluding mid-IR photometry (from SMILES) affects the estimated photo-$z$. 
Overall, both results are consistent with the JADES photo-$z$ \citep{JADES_Rieke2023ApJS..269...16R}. 
However, when including MIRI observations (labeled as NIR+MIR in Figure \ref{fig:eazy}), we find that the scatter $\sigma_\mathrm{NMAD}$ (defined as $1.48 \times \mathrm{median}\{|\Delta z|/(1+z_\mathrm{spec})\}$) is 0.0194, and the catastrophic outlier fraction $\eta$ (defined as the fraction of sources with $|\Delta z|/(1+z_\mathrm{spec}) > 0.15$) is 6.2\%, which is slightly worse than the results when only the NIRCam data are used (labeled as NIR-only, $\sigma_\mathrm{NMAD}$ = 0.0169, $\eta$ = 3.6\%). 
We find that the larger discrepancy in the NIR+MIR setup is due to the relatively high uncertainties of MIRI at longer wavelengths and incomplete mid-IR templates in EAZY, resulting in a poorer performance of EAZY during zero-point calibration. 
By carefully examining the eight EAZY photo-z outliers that were introduced with the inclusion of MIRI photometry, we notice that four of them lacked reliable detection at $>10$ $\mu$m, which had either low SNR ($<3$), or only upper limits, thus confusing the fitting. 
The other 4 showed apparent excess in the longer MIRI band, suggesting a possible lack of proper AGN templates for these galaxies.
In all subsequent calculations, we adopt redshift from EAZY photo-$z$ with NIR-only data and spec-$z$ from NIRSpec when available, to minimize uncertainty and improve constraints on SEDs and LFs.

\subsection{SED Fitting with CIGALE}
\label{sec:sed_fitting}
After obtaining the redshift prior from EAZY, we perform SED fitting with CIGALE v2022.1 and derive the physical properties of our galaxies. 
We utilize all available photometry of the galaxies in the catalog (i.e., optical to mid-IR bands) as input for CIGALE, where non-detection bands are treated as upper limits.
CIGALE provides multiple physical modules that can fit SEDs from X-ray to radio and generate comprehensive galaxy templates.
The specific modules and parameters used are based on \cite{Yang2023ApJ...950L...5Y} and \citetalias{Ling2024}, considering AGN emission and high-z galaxies. 
This configuration has performed well on JWST samples, but we have made slight optimizations based on CIGALE's Bayesian analysis, where the main change is to refine the grid of the star formation history module.
Table \ref{tab:cigale} shows the complete list of modules and parameters we used. 
Unlisted parameters remain the default of CIGALE.

We determine the galaxy type to be SF or AGN based on the CIGALE parameter frac$_{\rm AGN}$. 
Specifically, we use the convention \citep[e.g.,][]{Wang2020MNRAS.499.4068W, Chien2024MNRAS.532..719C, Hsieh2025}
\begin{align}
    {\rm frac_{AGN}} \equiv \frac{L_{\rm AGN}}{L_{\rm AGN}+L_{\rm galaxy}} \ge 0.2
\end{align}
as the criteria for classifying an AGN, i.e., the AGN luminosity ($3-30$ $\mu$m) should contribute more than 20\% of the total galaxy luminosity.
A total of 19\% (534) of galaxies are identified as AGN hosts in our sample.
As a rough check, we also cross-match our AGN sample against the X-ray data from Chandra Deep Field-South Survey \citep{Luo2017ApJS..228....2L} and radio data from VLA 1.4GHz ECDF-S survey \citep{Miller2013ApJS..205...13M}. 
Results show that only 37 of our AGN sources have X-ray counterparts, while 8 have both radio and X-ray counterparts. 
Interestingly, 5 of the X-ray detected sources were previously classified as SF galaxies in \cite{Luo2017ApJS..228....2L}. 
Their SEDs exhibit composite features, suggesting they are heavily obscured AGN that can only be effectively identified through mid-IR selection.

Still, we need to emphasise that the number of these counterparts should be considered as a lower limit because of the difference in survey coverage and depth. 
We remind that many AGNs observed by JWST are already considered X-ray faint \citep[e.g.,][]{Yang2023ApJ...950L...5Y, Chien2024MNRAS.532..719C, Maiolino2025MNRAS.538.1921M}, and may be heavily obscured AGNs detectable only in the IR band. 
Furthermore, only about $\sim15-20\%$ of AGNs are radio loud \citep{Urry1995PASP..107..803U}, so these results should not be conclusive on the completeness of our AGN sample.

We caution that due to the unavailability of FIR observations at similar resolutions (such as Herschel or {\it AKARI}), our SED curves, especially the FIR peak (dust-dominated), rely on information from shorter MIR wavelengths. 
The potential impact on the derived $L_{\rm IR}$ luminosity and related issues for higher-$z$ galaxies have been thoroughly discussed in \citetalias{Ling2024}; here, we only reiterate that for SF galaxies, the empirical relationship between $L_\mathrm{MIR}$ and $L_{\rm IR}$ allows us to use MIR observations as a proxy for the infrared SED of galaxies \citep[e.g.,][]{Caputi2007, Goto2011MNRAS.410..573G, Lin2024}, and should guarantee the validity of our SED fitting.
The agreement between our final LFs and ALMA-based LFs should further support this assumption, as presented in Section \ref{sec:TIR_LF}.

\begin{deluxetable*}{lll}
\tablewidth{0pt}
\tablecaption{CIGALE modules and parameters \label{tab:cigale}}
\tablehead{
\begin{tabular}[c]{@{}l@{}}\textbf{Module}\end{tabular} &
          \textbf{Parameters} &
          \textbf{Values}
}
\startdata
&
          Stellar e-folding time {[}Gyr{]} &
          0.5, 1, 2, 4 \\
        \multirow{-2}{*}{\begin{tabular}[c]{@{}l@{}}Star formation history\\ \texttt{sfhdelayed}\end{tabular}} &
          Stellar age {[}Gyr{]} &
          0.2, 0.5, 1, 2, 4 \\\hline
         &
          Initial mass function &
          \cite{Chabrier2003PASP..115..763C} \\
        \multirow{-2}{*}{\begin{tabular}[c]{@{}l@{}}Simple Stellar population\\ \texttt{bc03}\end{tabular}} &
          Metallicity &
          0.02 \\\hline
         &
          Ionisation parameter {[}log{]} &
          $-2.0$ \\
        \multirow{-2}{*}{\begin{tabular}[c]{@{}l@{}}Nebular emission\\ \texttt{nebular}\end{tabular}} &
          Gas metallicity &
          0.2 \\\hline
         &
          V-band attenuation in the interstellar medium ($A_V^{\rm ISM}$) &
          0.01, 0.02, 0.04, 0.08, 0.16, 0.32, 0.63, 1.3, 2.5, 5, 10 \\
         &
          $A_V^{\rm ISM}$ / ($A_V^{\rm BC}+A_V^{\rm ISM}$) &
          0.44 \\
         &
          Power law slope of the attenuation in the ISM &
          $-0.9$, $-0.7$, $-0.5$ \\
        \multirow{-4}{*}{\begin{tabular}[c]{@{}l@{}}Dust attenuation\\ \texttt{dustatt\_modified\_CF00}\end{tabular}} &
          Power law slope of the attenuation in the birth clouds &
          $-1.3$, $-1.0$, $-0.7$ \\\hline
         &
          PAH mass fraction &
          0.47, 2.5, 7.32 \\
         &
          Minimum radiation field &
          10, 15, 20 \\
        \multirow{-3}{*}{\begin{tabular}[c]{@{}l@{}}Galactic dust emission\\ \texttt{dl2014}\end{tabular}} &
          Fraction of PDR emission &
          0.01, 0.02, 0.05, 0.1, 0.2, 0.5 \\\hline
         &
          Average edge-on optical depth at 9.7 $\mu$m &
          3, 5, 7, 9, 11 \\
         &
          Viewing angle &
          $70^{\circ}$ \\
         &
          AGN contribution to IR luminosity &
          0, 0.01, 0.03, 0.05, 0.1, 0.2, 0.3, 0.5, 0.75, 0.9 \\
        \multirow{-4}{*}{\begin{tabular}[c]{@{}l@{}}AGN (UV-to-IR) emission\\ \texttt{skirtor2016}\end{tabular}} &
          Wavelength range where frac$_{\rm AGN}$ is defined &
          $3-30$ $\mu$m \\\hline
        \begin{tabular}[c]{@{}l@{}}Redshift+IGM\\ \texttt{redshift}\end{tabular} &
          redshift &
          EAZY output
\enddata
\end{deluxetable*}

\subsection{Quality Control}
\label{sec:qc}
\begin{figure}
    \centering
    \includegraphics[width=.5\textwidth] {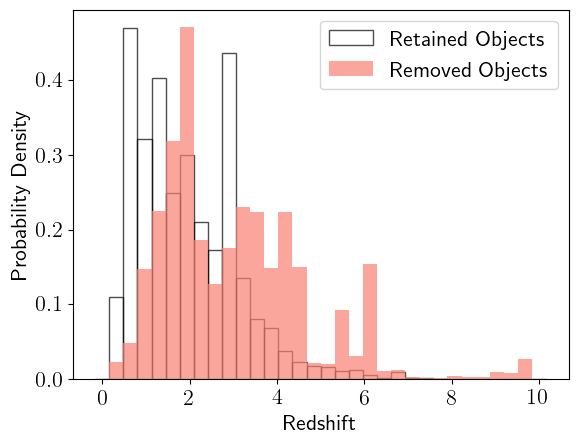}
    \caption{
    The stacked redshift (photo-$z$) probability distribution function from EAZY, normalized to 1.
    }
    \label{fig:z_pdf}
\end{figure}

Here, we describe how we handle unreliable or suspicious results in the fitting. 
For photo-$z$, we evaluate the redshift PDF, $\ln P(z)$, of each source. 
If there is no peak in the redshift PDF, or if the ratio between the likelihood of the secondary peak and the primary peak is greater than 50\% ($P_{2\mathrm{nd}}/P_\mathrm{main} > 50\%$), we mark the redshift estimate as unreliable and remove the source from the catalog.
In Figure \ref{fig:z_pdf}, we present a comparison between the stacked redshift PDF (normalized to 1) from EAZY for our final, retained galaxy sample and for those galaxies that were removed from the analysis. 
The overall distributions are consistent, while we find a shift in the PDF of the removed sample at $z\sim2$ and $z\sim4$. 
This feature is a direct consequence of our selection criteria, as objects with bad photo-z are often those with bimodal PDFs, where the code cannot reliably distinguish the low-redshift Balmer break from the high-redshift Lyman break. 
However, we remind that such deviation is exaggerated in the normalized PDF, given that removed objects are only 2.3\% of the sample.

\begin{figure}
    \centering
    \includegraphics[width=.5\textwidth] {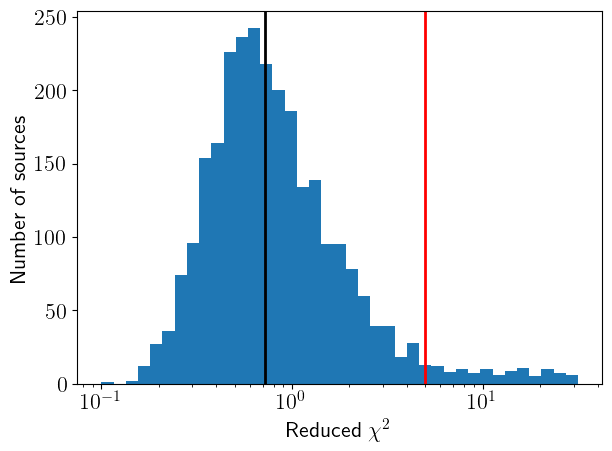}
    \caption{
    Distribution of the reduced $\chi^2$ values from the CIGALE SED fitting for our full sample. 
    The black vertical line indicates the median and the red vertical line marks our criteria of $\chi^2 = 5$.
    Above this, we consider the SED fit to be unreliable.}
    \label{fig:cigale}
\end{figure}

For SED, we require the reduced $\chi^2$ of the CIGALE fit to be less than 5, corresponding to 2$\sigma$ off the median.
The distribution and criteria of $\chi^2$ are shown in Figure \ref{fig:cigale}, where the median is 0.73.
Among the bad SED fits we removed, we find sources with $\chi^2 > 10$ (68\% of total removal) generally lack photometry in part of the HST or NIRCam bands, or only have upper limits, causing more significant discrepancies in the fitting results. Sources with $10 > \chi^2 > 5$ do not show apparent deficiencies, but only slightly poorer results.

\begin{figure}
    \centering
    \includegraphics[width=.5\textwidth] {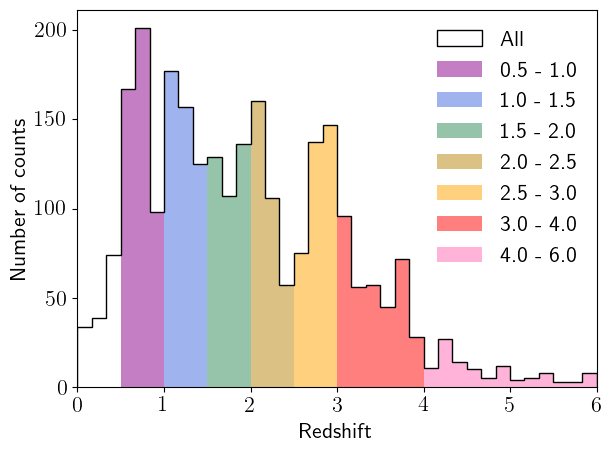}
    \caption{
    Photometric redshift distribution of the final 2607 sources. 
    Filled-colored histograms represent the distribution within each of the seven redshift bins used for the luminosity function analysis. 
    }
    \label{fig:zbin}
\end{figure}

In total, our selection removes 203 ($\sim$7.2\%) sources, including:
\begin{itemize}
    \item Bad photo-$z$ (EAZY): 64
    \item Bad SED fit (CIGALE): 139
\end{itemize}
To compensate for the removal, we multiply by a factor of 1.1 when calculating the LF (see next section), assuming that the redshift distribution of these removed galaxies is the same as the remaining ones.
Our final sample contains 2610 reliable sources.
Figure \ref{fig:zbin} shows their photometric redshift distribution and the corresponding 7 redshift bins we decided for LF: $z = (0.5-1.0), (1.0-1.5), (1.5-2.0), (2.0-2.5), (2.5-3.0), (3.0-4.0), (4.0-6.0)$.

\section{Luminosity functions}
\label{sec:LF}
\subsection{Computing LF}
\label{sec:compute_LF}
The process of obtaining the luminosity function is the same as in \citetalias{Ling2024}. 
Here, we explain how the final luminosity function is derived through these steps. 
We start from the best-fit SED and fluxes of galaxies in the observed frame from CIGALE.
To compute the rest-frame luminosity (for each filter, i.e., monochromatic luminosity), it is necessary to account for the SED redshift effect on the filter response curves, known as the K-correction. 
In the AB magnitude system, its overall effect can be described by the following equation:
 \begin{align}
    K(z) = \frac{1}{1+z}\frac{\int d\lambda_{\rm o} \;\lambda_{\rm o} L_\lambda\left(\frac{\lambda_{\rm o}}{1+z}\right)S(\lambda_{\rm o})}{\int d\lambda_{\rm e} \;\lambda_{\rm e} L_\lambda\left(\lambda_{\rm e}\right)S(\lambda_{\rm e})}
\end{align}
where $\lambda_{\rm o}$ and $\lambda_{\rm e}$ are the filter wavelength in the observed and rest-frame, respectively, $S\left(\lambda\right)$ is the transmission curve for the specific filter, and $L_\lambda$ is the luminosity density per unit wavelength. In practice, the correction is done by deriving the rest-frame flux $F_\lambda(\lambda_{\rm o})$, where we shift the SED to the rest-frame and re-integrate it against the given filter $S\left(\lambda\right)$. 
Then, with a rest-frame flux and $z$, we can determine the rest-frame luminosity $L_\lambda$:
\begin{align}
    L_\lambda(\lambda_{\rm e}) &= \frac{4\pi D_{\rm L}(z)^2}{1+z} F_\lambda(\lambda_{\rm o})
\end{align}
with $\lambda_{\rm e} = \lambda_{\rm o}/(1+z)$ and the luminosity distance at redshift $z$ ($D_{\rm L}$).
For $L_{\rm IR}$ ($8-1000$ $\mu$m), a specific correction to the filter flux is not needed. 
Instead, we simply convert the SED to the rest-frame and integrate directly, treating it as a flat-response filter over $8-1000$ $\mu$m.

To ensure the completeness of our LF, we introduce luminosity limits for each redshift bin used.
For the monochromatic LFs, the limiting luminosity is estimated independently for each band, accounting for their different depths.
For the IR LF, as the sample is selected in the F560W band, we use the F560W completeness to determine the limiting IR luminosity. 
Similar to \citetalias{Ling2024}, we calculate the limiting luminosity at each redshift by assuming two galaxy SED templates, NGC6090 (for SF galaxies) and Sey2 (for AGN host galaxies) from the SWIRE Template Library \citep{Polletta2007}, when they are at the 80\% completeness flux.

\begin{figure*}
    \centering
    \includegraphics[width=\textwidth] {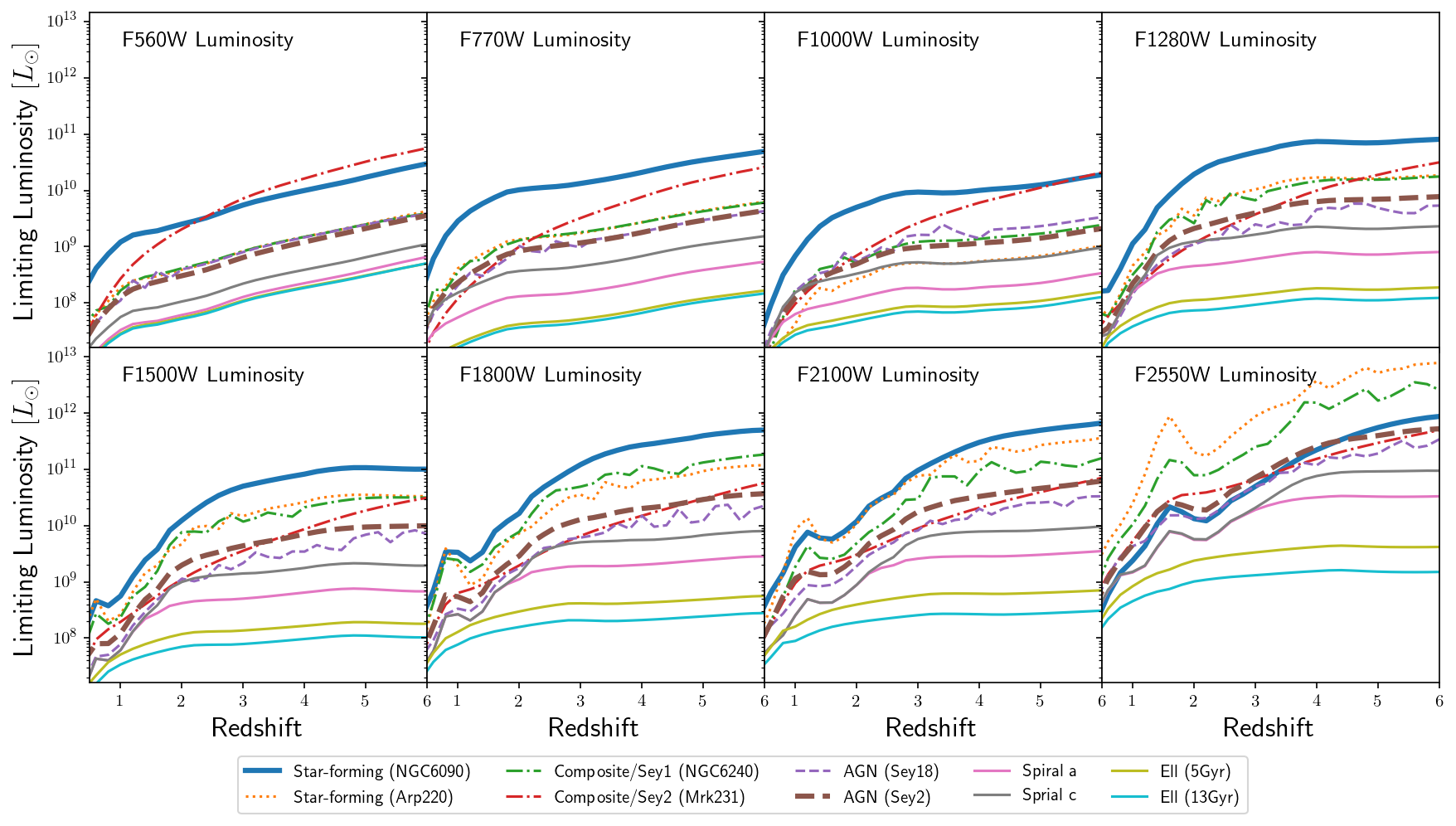}
    \caption{
    Limiting monochromatic luminosity as a function of redshift for a range of galaxy SED templates. 
    Each panel shows the luminosity limit for a specific MIRI band, derived from its 80\% completeness flux limit. 
    Different line styles and colors correspond to various templates, including star-forming galaxies (NGC6090, Arp220), composite galaxies (NGC6240, Mrk231), AGNs (Seyfert 1.8, 2), and quiescent populations (spirals and ellipticals). The thick solid blue line (NGC6090) represents our adopted template for star-forming galaxies, and the thick dashed brown line (Seyfert 2) is used for our AGN population.
    }
    \label{fig:lum_limit_full}
\end{figure*}

The selection of these two galaxy templates follows \citep{Yang2023ApJ...950L...5Y}, which suggests they resemble the median SED of the JWST MIRI-selected galaxies. 
We have verified the robustness of these adopted SED templates. 
For SF galaxies, we compared the limiting luminosities derived from various templates and found that NGC6090 provides one of the most conservative (highest) luminosity limits across the MIRI bands and redshift range considered, confirming it is a suitable choice (Figure \ref{fig:lum_limit_full}).
For the AGN population, the Seyfert 2 template was chosen because it best represents the MIRI-selected AGN sample \citep{Yang2023ApJ...950L...5Y}. 
While other AGN templates could be chosen, the stable faint-end slope of our AGN LF demonstrates that our primary results are not sensitive to a more conservative luminosity limit.

\begin{figure}
    \centering
    \includegraphics[width=.5\textwidth] {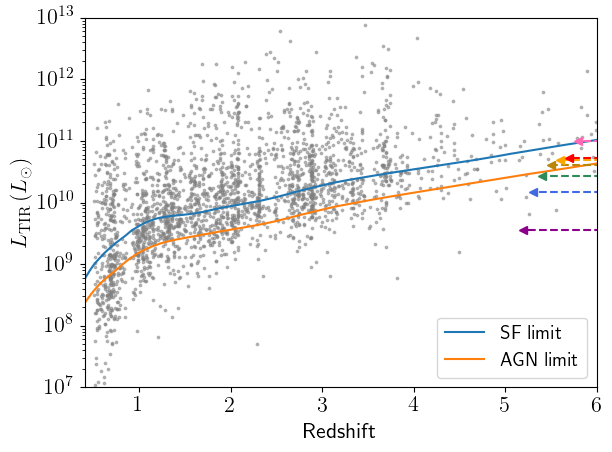}
    \caption{
    Limiting $L_{\rm IR}$ luminosity as a function of redshift for the two adopted SED templates, assuming the 80\% completeness limit of F560W.
    The colored (see Figure \ref{fig:zbin}) arrows indicate the adopted limiting luminosity for each of the seven redshift bins.
    }
    \label{fig:lum_limit}
\end{figure}

We thus have two "luminosity limit functions" (for SF and AGN galaxies, respectively) to cut off our LFs.
Figure \ref{fig:lum_limit} shows an example of the luminosity limit function for the $L_{\rm IR}$ LF (using the F560W band limit).
The maximum limiting luminosity in each redshift interval (indicated by colored arrows in the figure, considering both SED types) is the luminosity limit for that redshift. 
Galaxies with luminosities below this value are considered incomplete and are not included in the analysis.

The LF is derived using the classical 1/$V_\mathrm{max}$ method \citep{Schmidt1968ApJ...151..393S}, which addresses the volume incompleteness of flux-limited samples:
\begin{align}
    \phi (L) = \frac{1}{\Delta \log L} \sum_i \frac{1}{V_{\mathrm{max}, i}}w_{i, \nu}
\end{align}
where we adopt $\Delta \log L = 0.5$ dex as the luminosity bin width, and $V_{\mathrm{max}, i} = V(z_\mathrm{max}) - V(z_i)$ is the maximum comoving volume corresponding to galaxy $i$. 
The $z_\mathrm{max}$ is calculated by the same method used in \citetalias{Ling2024}. 
$w_{i, \nu}$ is the correction factor:
\begin{align}
    w_{i, \nu} = \frac{4\pi \ \mathrm{sr}}{\mathrm{area}} \frac{\mathrm{Compensation}}{\mathrm{completeness}_{i, \nu}(F_{i, \nu})}
\end{align}
which considers the completeness of a given galaxy $i$ at flux $F$ in band $\nu$, and an overall compensation of $\sim$1.10 (10 percent) to correct for the galaxies removed in Section \ref{sec:qc}. 
The uncertainty of our LF includes the Poisson error and the photo-$z$ distribution. 
To estimate the error introduced by EAZY photo-$z$, we sample redshifts of galaxies from the PDF 100 times. 
For each sample, we keep the best-fit SED obtained from the original CIGALE fitting but re-calculate the galaxy luminosities using the newly sampled redshift. The distribution of these 100 luminosity values provides an estimate of the uncertainty due to the photometric redshift error.
Then, we analyze the distribution of luminosity bins in new LFs, adding a Poisson error to each luminosity bin as the final uncertainty.

\subsection{Monochromatic LFs}
\label{s:monoLF}
\begin{figure*}
    \centering
    \includegraphics[width=\textwidth] {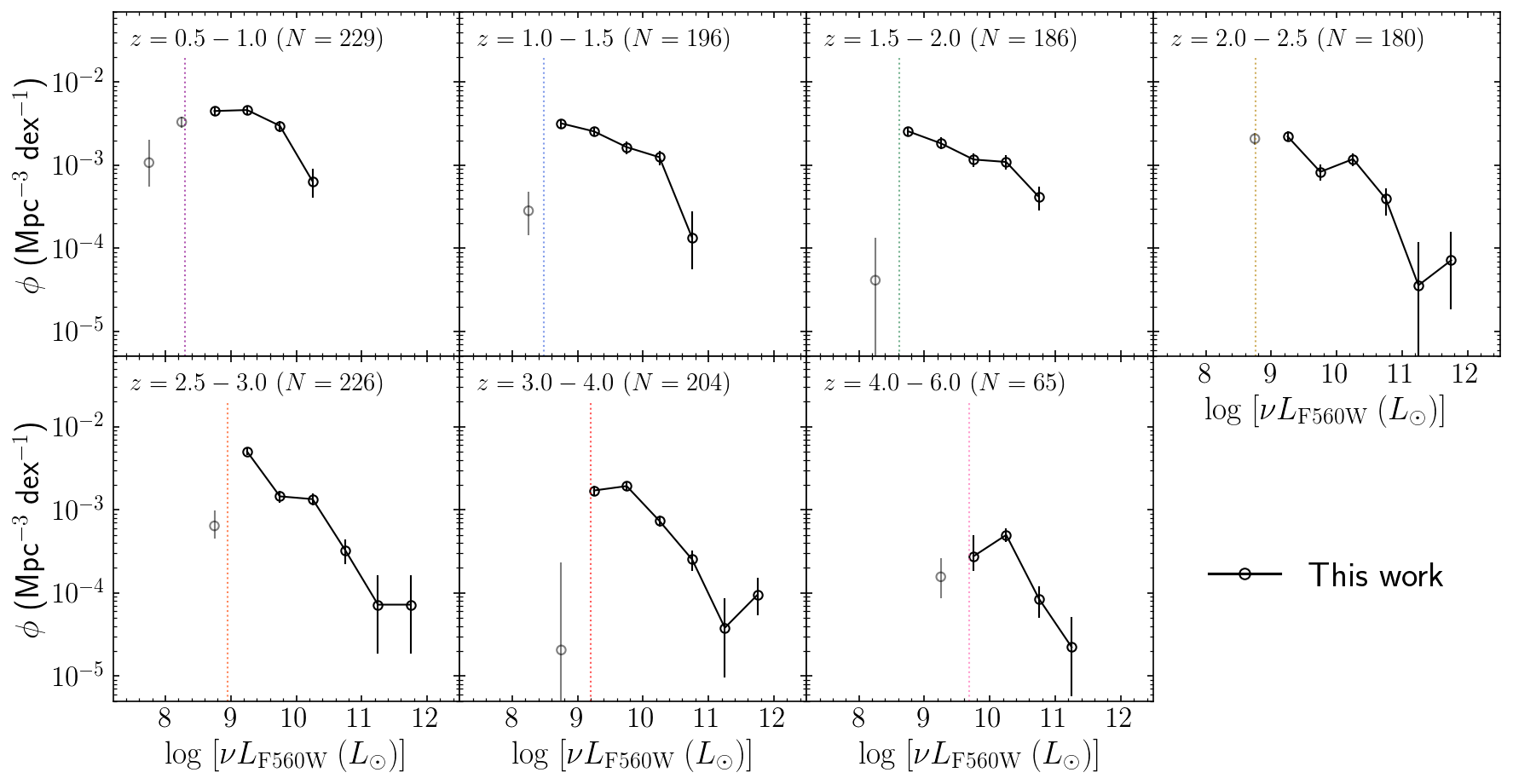}
    \caption{
     Rest-frame 5.6 $\mu$m (F560W) monochromatic luminosity ($\nu L_\nu$) functions (open black circle). 
     Vertical dashed lines indicate the luminosity at which the completeness drops to 80\%.
     The data points below the limits are shown in gray and are not connected.
     The number of galaxies in the redshift bin is indicated at each panel's top.
    }
    \label{fig:LF_560}
\end{figure*}

\begin{figure*}
    \centering
    \includegraphics[width=\textwidth] {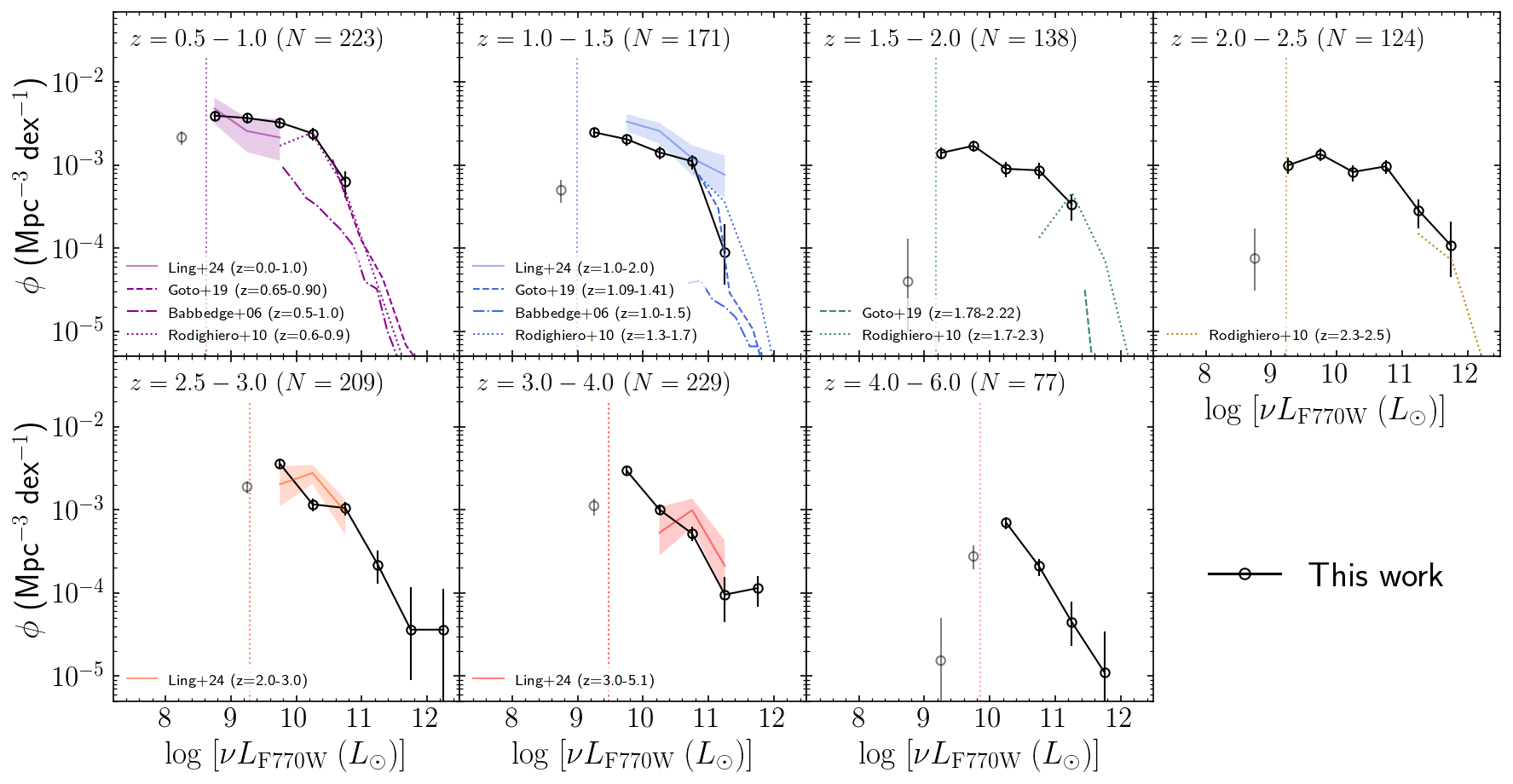}
    \caption{
    7.7 $\mu$m (F770W) luminosity functions. 
    LFs from \protect\citet[dot-dashed line]{Babbedge2006MNRAS.370.1159B}, \protect\citet[dotted line]{Rodighiero2010A&A...515A...8R}, \protect\citet[dashed line]{Goto2019PASJ...71...30G}, and previous JWST work, \protect\citet[solid line]{Ling2024}, are also plotted.
    }
    \label{fig:LF_770}
\end{figure*}

\begin{figure*}
    \centering
    \includegraphics[width=\textwidth] {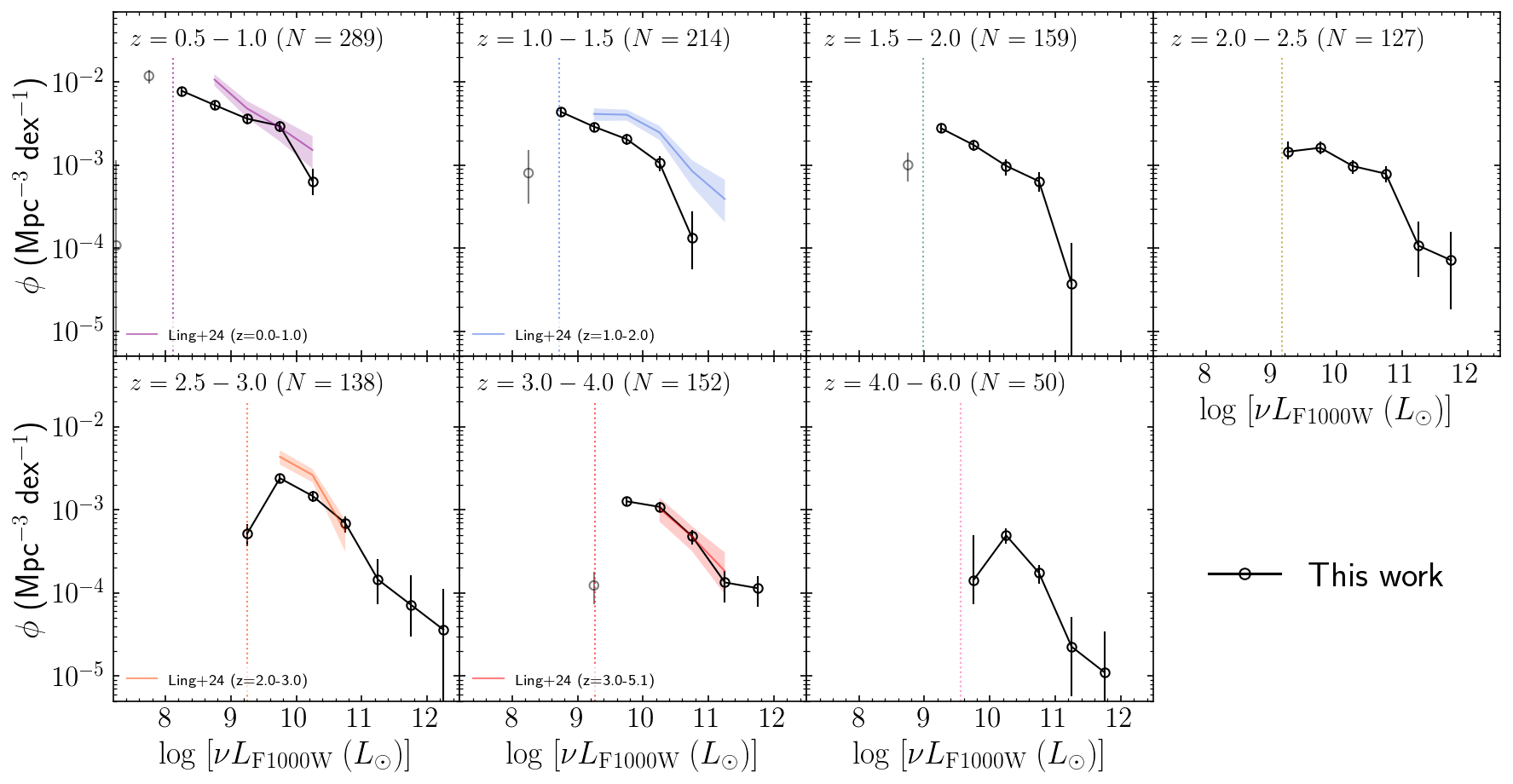}
    \caption{
    10 $\mu$m (F1000W) luminosity functions. 
    LF from previous JWST work, \protect\citet[solid line]{Ling2024}, is also plotted.
    }
    \label{fig:LF_1000}
\end{figure*}

\begin{figure*}
    \centering
    \includegraphics[width=\textwidth] {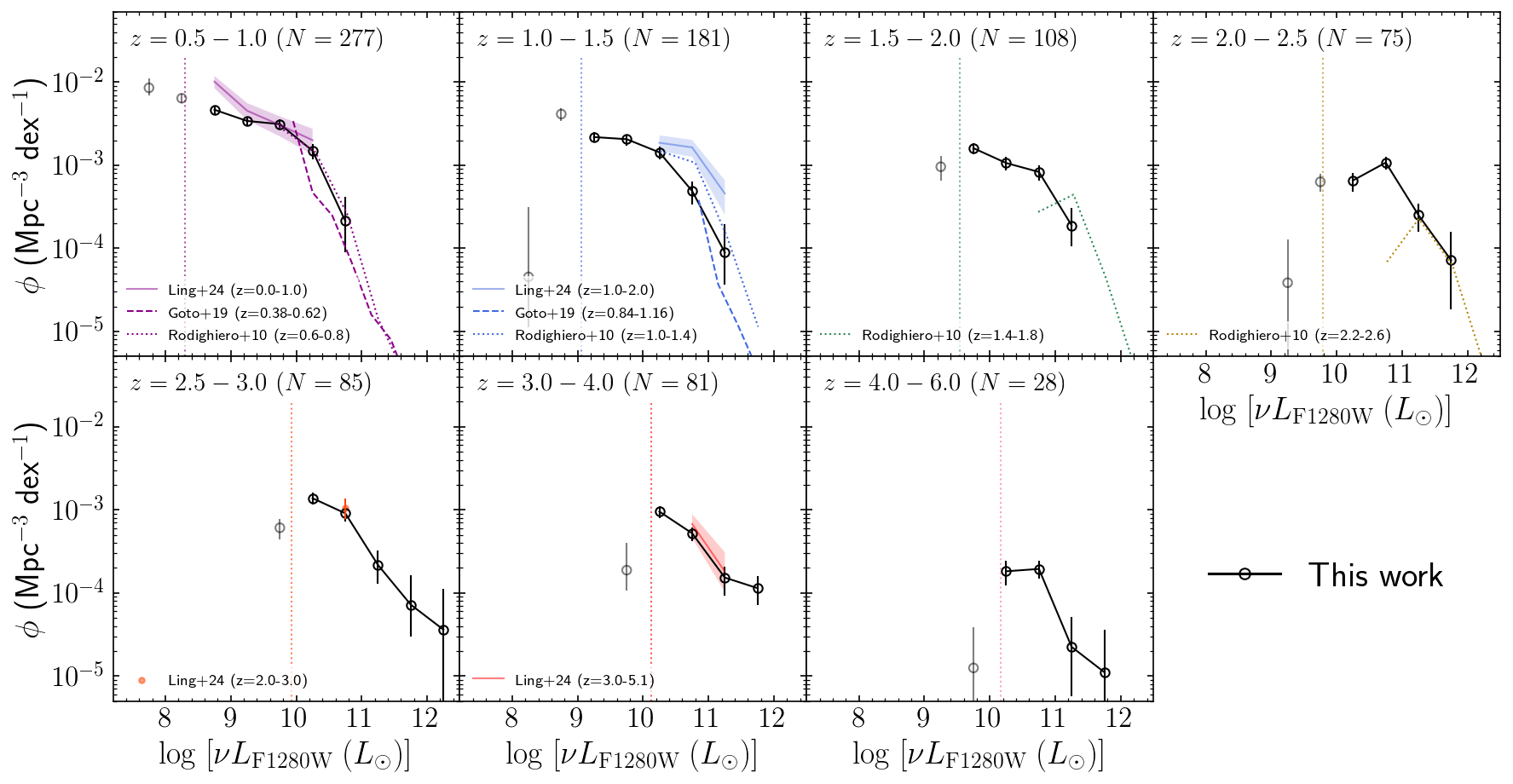}
    \caption{
    12.8 $\mu$m (F1280W) luminosity functions. 
    LFs from \protect\citet[dotted line]{Rodighiero2010A&A...515A...8R}, \protect\citet[dashed line]{Goto2019PASJ...71...30G}, and \protect\citet[solid line]{Ling2024} are also plotted.
    }
    \label{fig:LF_1280}
\end{figure*}

\begin{figure*}
    \centering
    \includegraphics[width=\textwidth] {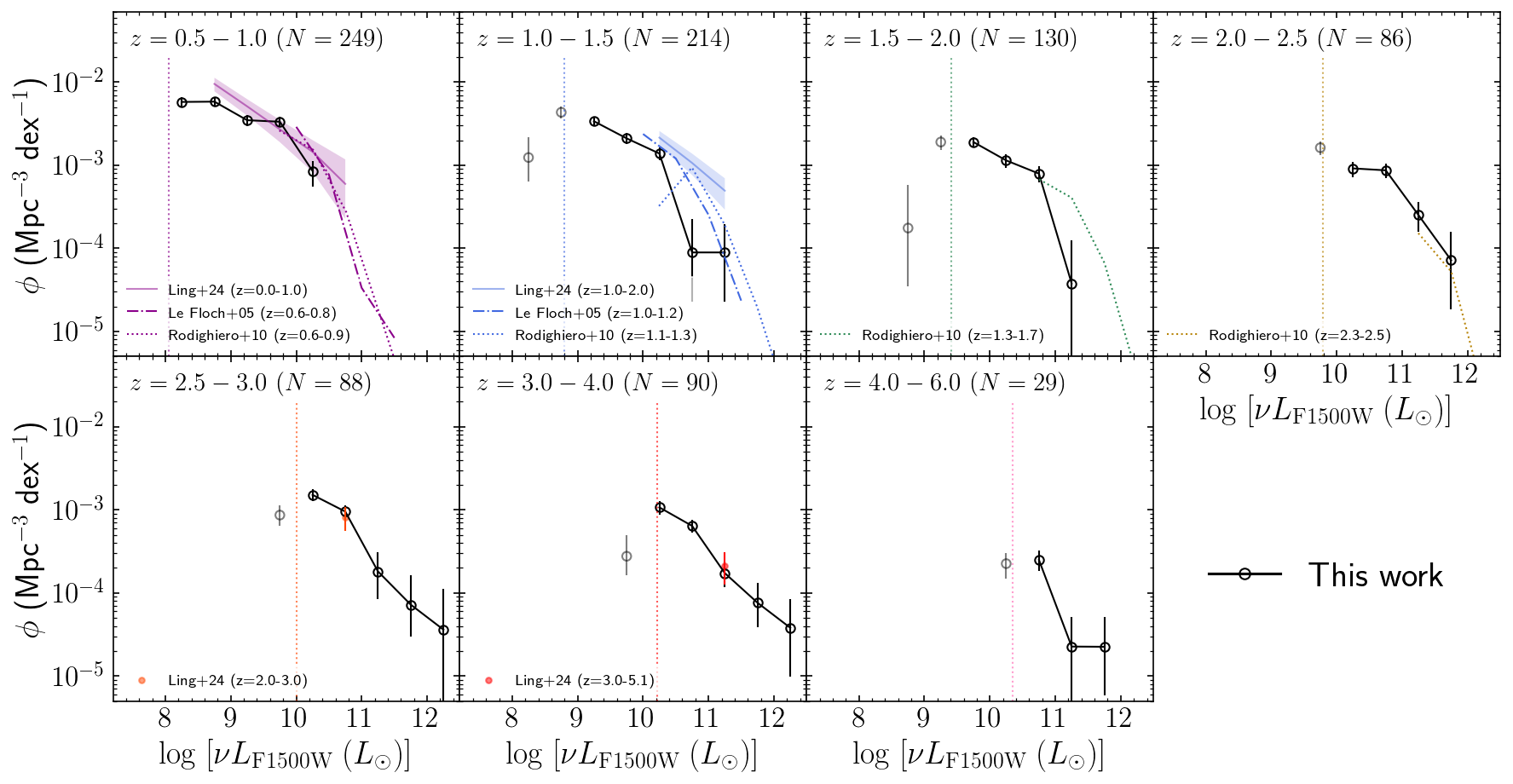}
    \caption{
    15 $\mu$m (F1500W) luminosity functions. 
    LFs from \protect\citet[dot-dashed line]{LeFloch2005ApJ...632..169L}, \protect\citet[dotted line]{Rodighiero2010A&A...515A...8R}, and \protect\citet[solid line]{Ling2024} are also plotted.
    }
    \label{fig:LF_1500}
\end{figure*}

\begin{figure*}
    \centering
    \includegraphics[width=\textwidth] {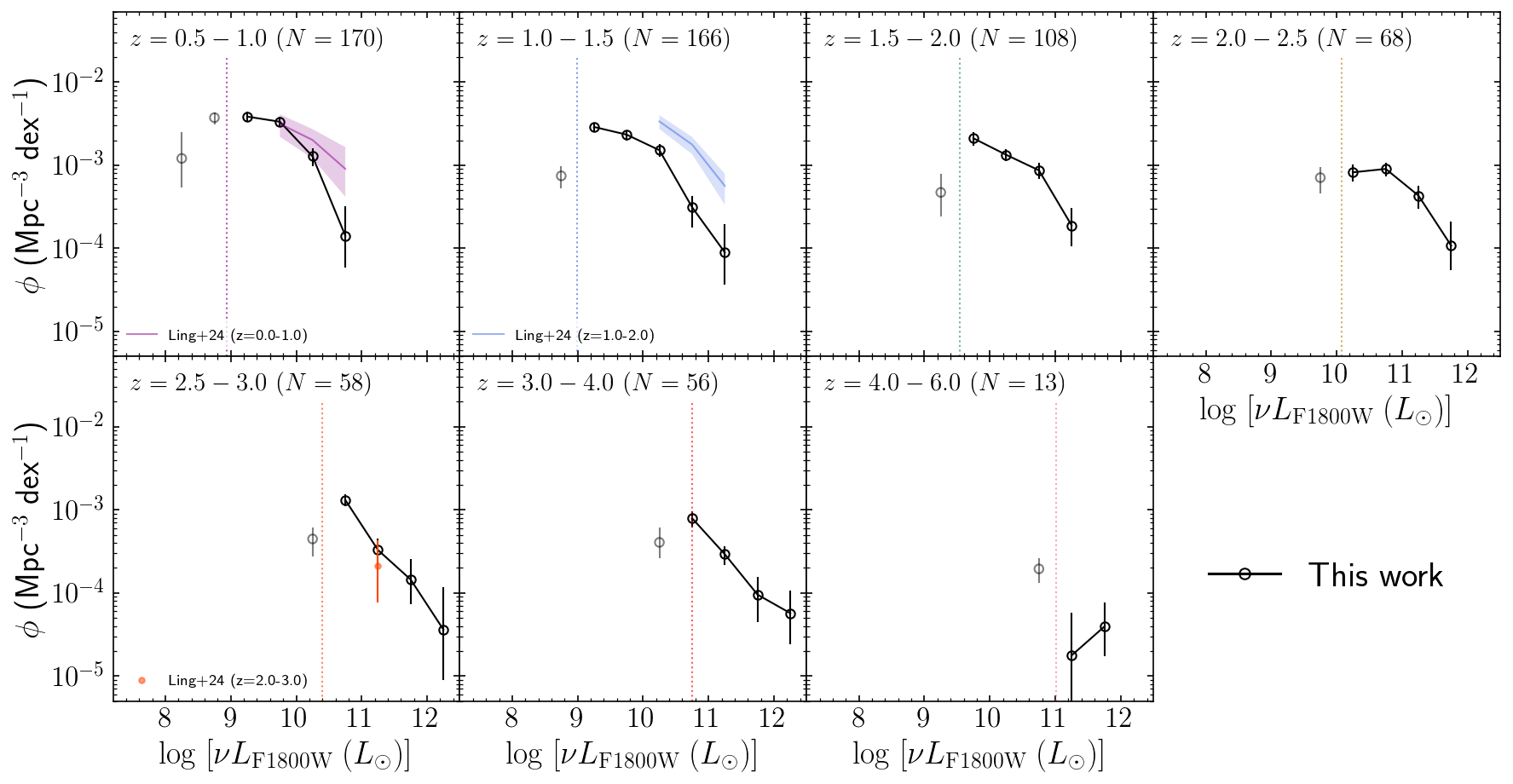}
    \caption{
    18 $\mu$m (F1800W) luminosity functions. 
    LF from \protect\citet[solid line]{Ling2024} is also plotted.
    }
    \label{fig:LF_1800}
\end{figure*}

\begin{figure*}
    \centering
    \includegraphics[width=\textwidth] {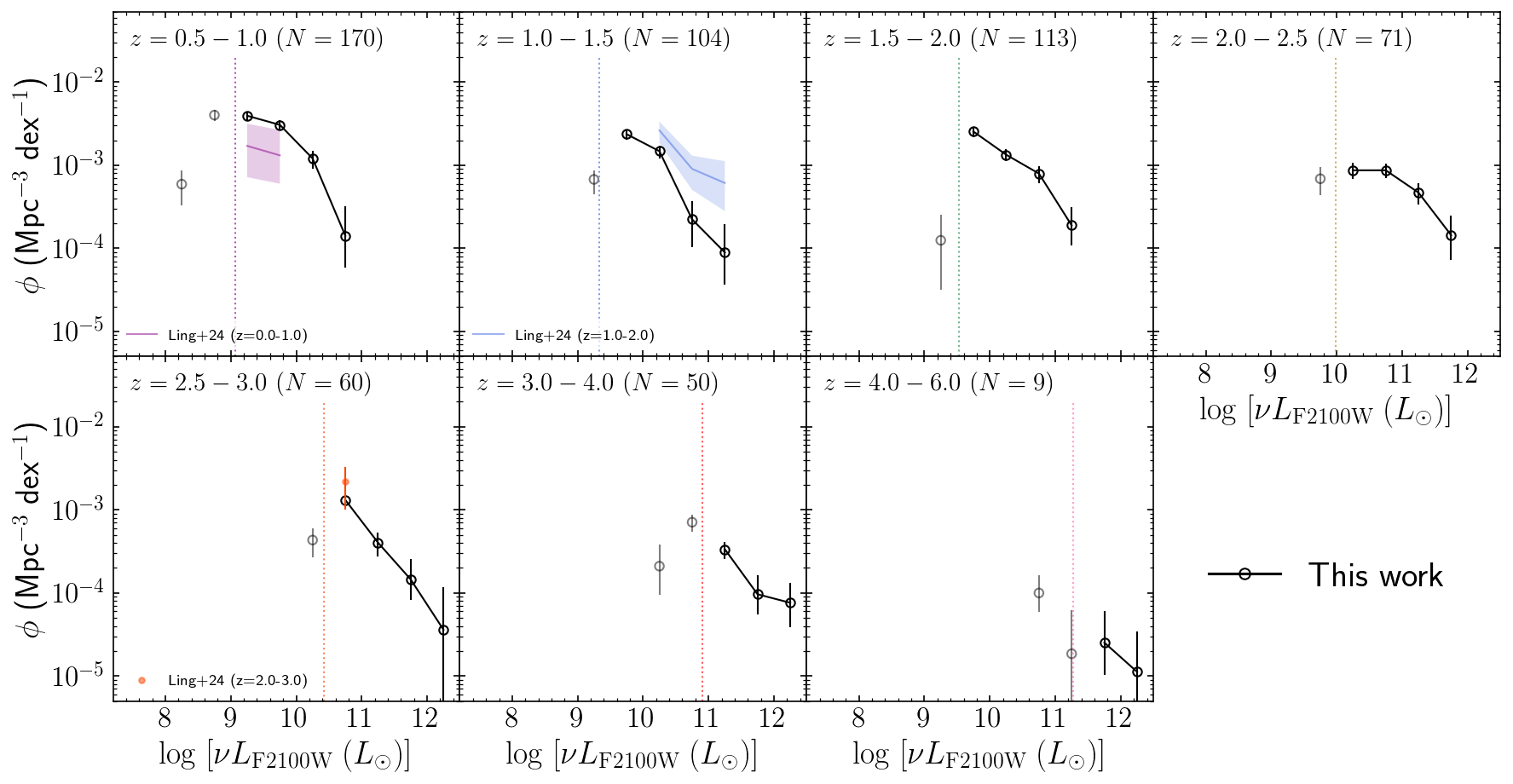}
    \caption{
    21 $\mu$m (F2100W) luminosity functions. 
    LF from \protect\citet[solid line]{Ling2024} is also plotted.
    }
    \label{fig:LF_2100}
\end{figure*}

\begin{figure*}
    \centering
    \includegraphics[width=\textwidth] {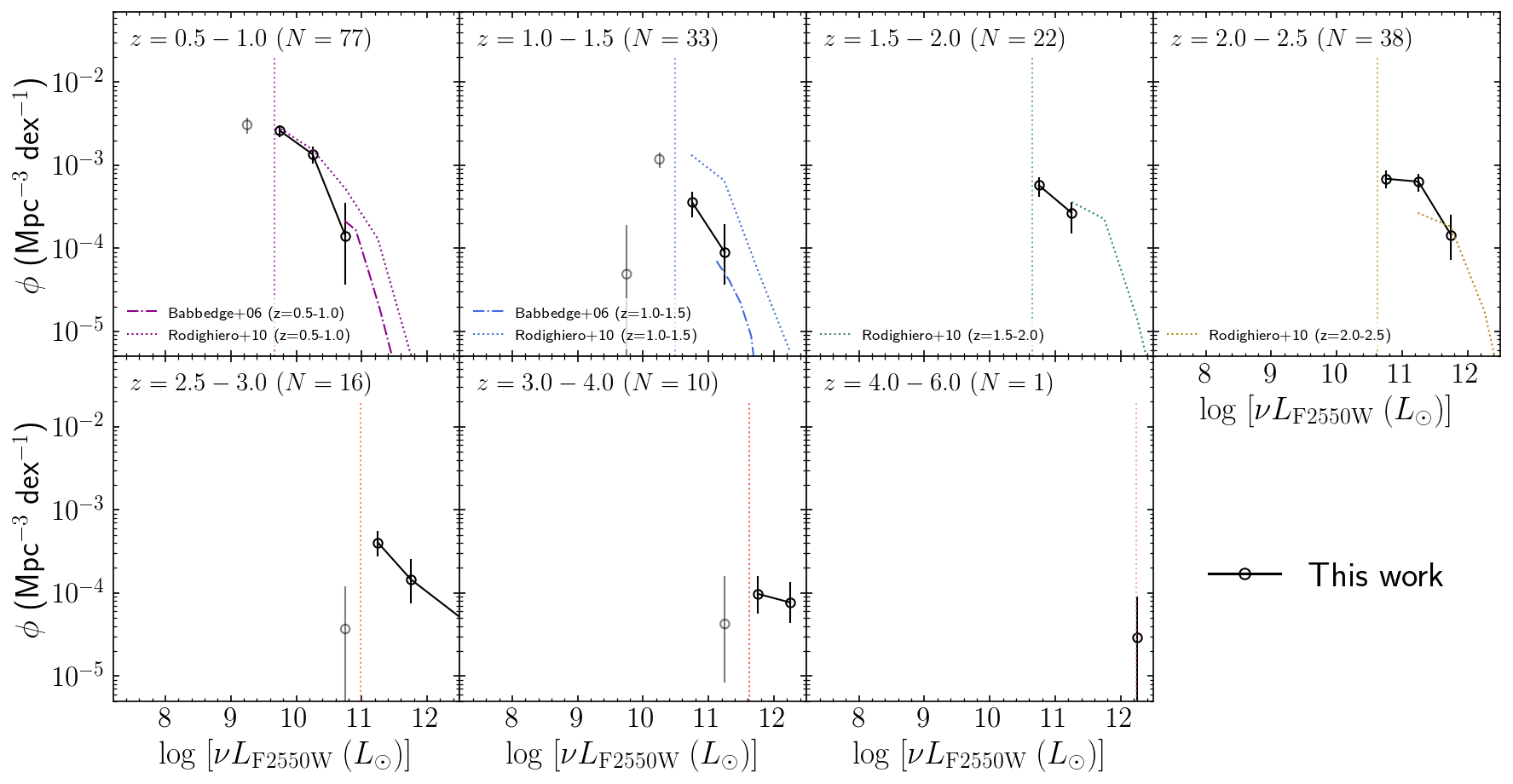}
    \caption{
    25.5 $\mu$m (F2550W) luminosity functions. 
    LFs from \protect\citet[dot-dashed line]{Babbedge2006MNRAS.370.1159B} and \protect\citet[dotted line]{Rodighiero2010A&A...515A...8R} are also plotted.
    }
    \label{fig:LF_2550}
\end{figure*}

Figures \ref{fig:LF_560} to \ref{fig:LF_2550} show our monochromatic luminosity functions derived from SMILES, corresponding to 5.6, 7.7, 10, 12.8, 15, 18, 21, and 25 $\mu$m, respectively. 
Representative MIR luminosity functions from the literature are provided if their redshift range is similar to ours.
At 7.7, 12.8, 15, and 25.5 $\mu$m, we have LFs from AKARI \citep{Goto2019PASJ...71...30G} and Spitzer \citep{LeFloch2005ApJ...632..169L, Babbedge2006MNRAS.370.1159B, Rodighiero2010A&A...515A...8R} for comparison. 
The JWST-based \citetalias{Ling2024} is available at all wavelengths except 5.6 and 25.5 $\mu$m.

Overall, the trend of our monochromatic LFs is similar to the previous \citetalias{Ling2024}, and connects to the faint end of the LFs from the last generation of infrared space telescopes. 
The main luminosity coverage is $\sim 10^9 - 10^{11} L_\odot$. 
Compared to \citetalias{Ling2024}, this work has several improvements:
benefiting from the larger survey area and sample size, as well as the excellent photometric processing of SMILES, we have significantly reduced the uncertainty of each luminosity bin while maintaining a limiting luminosity similar to or even fainter than \citetalias{Ling2024}. 
At $z=0.5-1.0$, the lowest reliable luminosity that can be reached is $L_* = 10^{8.25} L_\odot$ (F1000W and F1500W). 
At the same time, we are able to adopt finer redshift bins than \citetalias{Ling2024}.
The luminosity range of the JWST LF can now be further extended to both ends rather than just one or two points. 
In addition, this is also the first time to obtain 5.6 and 25.5 $\mu$m LFs with JWST MIRI (from F560W and F2550W bands).

We compare and discuss our monochromatic LFs at each wavelength as follows.
\begin{itemize}
    \item \textbf{5.5 and 7.7 $\mu$m}
    While there is no literature to compare, we find the trend of 5.5 $\mu$m LFs is very similar to 7.7 $\mu$m LF. 
    In each redshift bin, our 7.7 $\mu$m LFs are consistent with the 8 $\mu$m LFs of \cite{Rodighiero2010A&A...515A...8R} and \cite{Goto2019PASJ...71...30G}, connecting at $L_* \sim 10^{11} L_\odot$. 
    This also suggests that the faint end of \cite{Babbedge2006MNRAS.370.1159B} is significantly underestimated.
    \citetalias{Ling2024} is slightly higher than our measured value at $z=1-2$, but overlaps at other redshift bins. 
    Our LFs cover a wider luminosity range ($>1$ dex) after $z>2.5$, and no fluctuation was shown as in \citetalias{Ling2024}.
    Although they have adopted the completeness limit, \citetalias{Ling2024} is more affected by the small statistics at the faint end.
    
    \item \textbf{10 $\mu$m}
    At $z<3$, our LFs are lower than \citetalias{Ling2024}, although some still intersect. 
    This discrepancy can also be seen in the LFs at longer wavelengths but is less significant. 
    We attribute this to the following reasons. 
    First, the sample of \citetalias{Ling2024} is not selected based on filters, while this work adopts the F560W+F770W selection of \cite{SMILES_Alberts2024}. 
    As \citetalias{Ling2024} allows sources detected in any MIRI band, the loose standard may cause them to select relatively more sources at long wavelengths (see also their Table 5). 
    Second, considering that the CEERS field area used by \citetalias{Ling2024} is only $\sim 9$ arcmin$^2$, the possibility of being affected by cosmic variance is more significant ($\sim 20\%$, as in \citealt{Moster2011ApJ...731..113M}). 
    We find that adjusting our LF redshift range to \citetalias{Ling2024}'s does not reduce this discrepancy, indicating that it may stem more from the above systematics.
    
    \item \textbf{12.8 and 15 $\mu$m}
    The LF trend is roughly the same as at 7.7 $\mu$m, and we find a good agreement with \cite{Rodighiero2010A&A...515A...8R} and \cite{Goto2019PASJ...71...30G}. 
    However, the faint end is lower compared to \citetalias{Ling2024} at $z=0-1$, as seen in the above discussion.
    
    \item \textbf{18 and 21 $\mu$m}
    These two wavelengths only have the data points of \citetalias{Ling2024} for comparison. 
    Except for the fact that \citetalias{Ling2024} is higher than our LF at $z=1-2$ in 18 $\mu$m, where we have an apparent drop at the bright-end, likely due to insufficient samples. In other redshift bins, the error bars of the two touch. 
    In addition, this is the first time that we push the LF to $z>3$ at these wavelengths; \citetalias{Ling2024} is limited by the sample size and thus lacks these rare high-$z$ galaxies.
    
    \item \textbf{25.5 $\mu$m}
    Here we compare our result with the 24 $\mu$m LFs of \cite{Babbedge2006MNRAS.370.1159B} and \cite{Rodighiero2010A&A...515A...8R}. 
    As in the above comparison, our LFs also agree with \cite{Rodighiero2010A&A...515A...8R} and have advanced 0.5 dex to the faint end at $z=1.5-2.0$ and $z=2.0-2.5$, respectively. 
    Due to the poor 80\% completeness limit of the F2550W band (only 15 $\mu$Jy, compared to 80 $\mu$Jy of MIPS 24 $\mu$m) and the relationship of the galaxy SED shape, the luminosity limits of the two at low redshift ($z<1.5$) are the same. 
    Still, we have pushed the LF to $z=2.5-4$ for the first time, although there is only a small sample of galaxies in this redshift interval.
\end{itemize}


\subsection{Infrared LFs}
\label{sec:TIR_LF}
\begin{figure*}
    \centering
    \includegraphics[width=\textwidth] {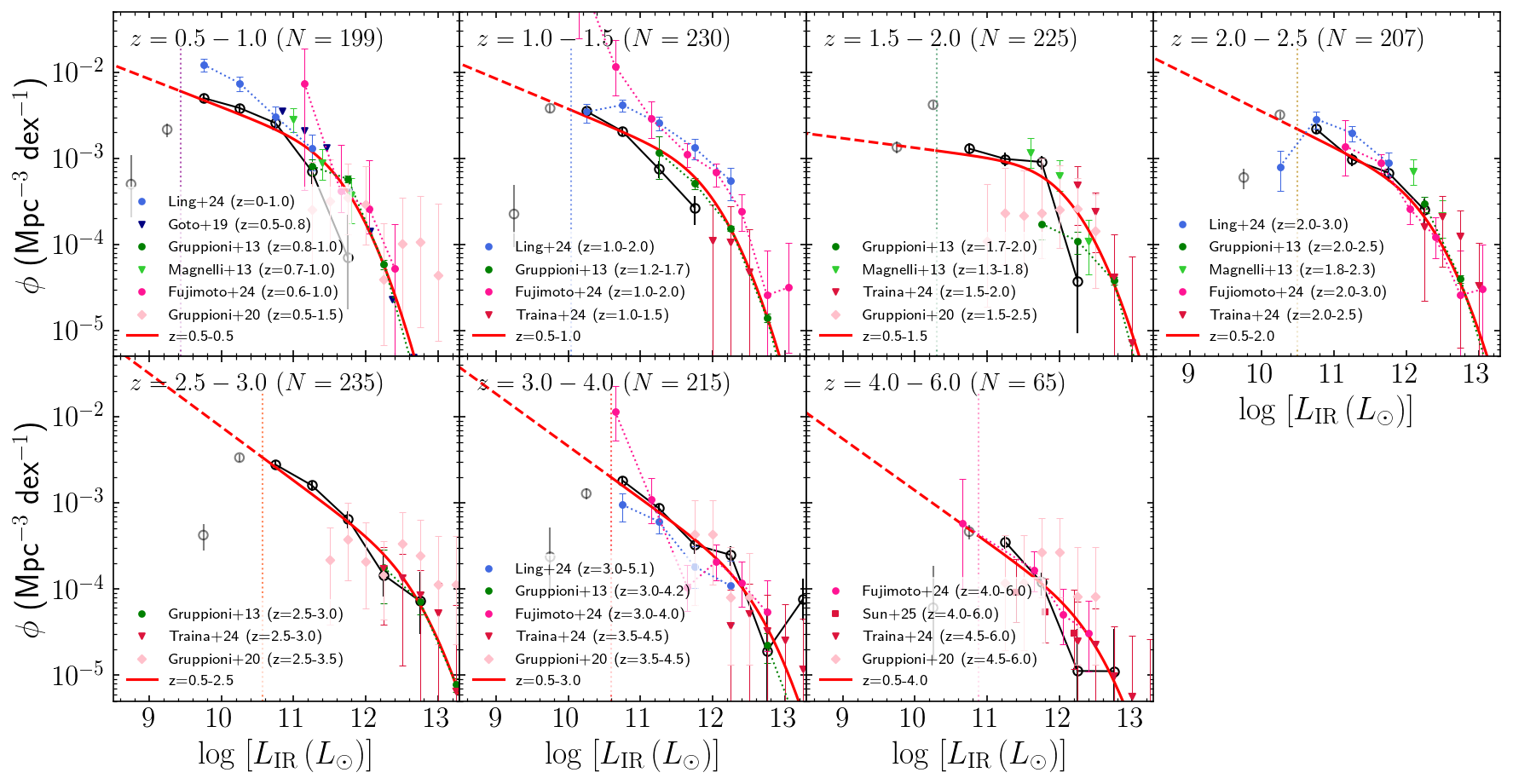}
    \caption{
    Infrared ($L_{\rm IR}$, $8-1000$ $\mu$m) luminosity functions (open black circle). 
    Vertical dashed lines indicate the luminosity at which the completeness drops to 80\%.
    The data points below the limits are shown in gray and are not connected.
    The number of galaxies in the redshift bin is indicated at each panel's top.
    The red solid line is the best fit to our data combined with \cite{Gruppioni2013MNRAS.432...23G}. 
    Fits below the luminosity limit are represented by dashed lines.
    LFs from \protect\citet[navy triangle]{Goto2019PASJ...71...30G}, \protect\citet[blue circle]{Ling2024}, \protect\citet[green circle]{Gruppioni2013MNRAS.432...23G}, \protect\citet[lime triangle]{Magnelli2013A&A...553A.132M}, \protect\citet[pink diamond]{Gruppioni2020A&A...643A...8G}, \protect\citet[magenta circle]{Fujimoto2024ApJS..275...36F}, \protect\citet[red triangle]{Traina2024A&A...681A.118T}, and \protect\citet[red square]{Sun2025ApJ...980...12S} are also plotted.
    }
    \label{fig:LF_TIR}
\end{figure*}

Figure \ref{fig:LF_TIR} shows our infrared ($L_{\rm IR}$, $8-1000$ $\mu$m) luminosity functions.
Our LFs mainly cover the luminosity range of $10^{10} - 10^{12} L_\odot$.
They are derived from our F560W-selected sample (refer to Figure \ref{fig:lum_limit} for the luminosity limit using F560W band completeness) in order to optimize the balance between depth, sample size, and physical information. We note that requiring detection at more or longer MIRI wavelengths would introduce a significant bias towards brighter sources, given their poorer sensitivity (e.g., Table \ref{tab:smiles}; F1800W is $10\times$ lower than F770W), and would exclude the crucial faint galaxy population needed to constrain the luminosity function. This would also reduce our sample size significantly, weakening the statistical power of our results.

For comparison, we provide representative $L_{\rm IR}$ LFs from previous works across different wavelengths: mid-IR based LFs from Spitzer \citep{LeFloch2005ApJ...632..169L}, AKARI \citep{Goto2019PASJ...71...30G}, and JWST \citetalias{Ling2024}; far-IR based LFs from Herschel \citep{Gruppioni2013MNRAS.432...23G, Magnelli2013A&A...553A.132M}; and sub-mm based LFs from ALMA studies \citep{Gruppioni2020A&A...643A...8G, Fujimoto2024ApJS..275...36F, Traina2024A&A...681A.118T, Sun2025ApJ...980...12S}. 

The $L_{\rm IR}$ LFs exhibit a similar evolutionary trend to the monochromatic LFs.
Our results are broadly consistent with the literature, particularly at the bright end ($L_{\rm IR} > 10^{11} L_\odot$), and significantly connect the pre-JWST LFs \citep[i.e.,][]{Gruppioni2013MNRAS.432...23G, Magnelli2013A&A...553A.132M} to fainter luminosities, reaching a limiting luminosity of $L_{\rm IR} \sim 10^{9} L_\odot$ at $z = 0.5 - 1.0$. 
This depth is comparable to, or slightly deeper (at $z>1$) than, previous JWST results (\citetalias{Ling2024}).
As with monochromatic LFs, the larger survey area and sample size of SMILES allow us to extract the faint end slopes at high-$z$ for the first time, which was not possible in \citetalias{Ling2024}.
While the general shape of our LFs agrees with \citetalias{Ling2024}, we note minor differences at lower redshift ($z<1.5$). As discussed in Section \ref{s:monoLF}, this discrepancy stems from variation in sample selection and/or the effects of cosmic variance. 

Studies utilizing ALMA sub-millimeter observations, which directly probe the bright peak of the rest-frame FIR dust emission, provide another independent constraint on the $L_{\rm IR}$ LF. 
It is crucial to compare our JWST LFs with these results, as ALMA observations can more completely capture the properties of the galaxy dust emission than LFs based solely on MIR extrapolations. 
Furthermore, as shown in Figure \ref{fig:LF_TIR}, the luminosity limits of ALMA LFs become comparable to, or even match (i.e., \citealp{Fujimoto2024ApJS..275...36F}), those reached with JWST at intermediate and higher redshifts ($z>1.5$).
We provide further discussion in Section \ref{sec:LF_fitting} on comparing the shape of this deep LF from \cite{Fujimoto2024ApJS..275...36F}.

In the overlapping luminosity range $L_{\rm IR} \approx 10^{11} - 10^{12.5} L_\odot$, we find good consistency between JWST LFs and results from all aforementioned ALMA works (\citealp{Gruppioni2020A&A...643A...8G, Fujimoto2024ApJS..275...36F, Traina2024A&A...681A.118T, Sun2025ApJ...980...12S}), except for the $z = 1.0-1.5$ bin where our overall number densities $\phi$ are slightly lower, though still consistent with \citetalias{Ling2024}. 
Our data points generally fall within the larger uncertainties of \cite{Gruppioni2020A&A...643A...8G}, connect smoothly to the bright end probed by \cite{Traina2024A&A...681A.118T}, and share a similar trend with \cite{Fujimoto2024ApJS..275...36F}. 
Our results at $z = 4.0-6.0$ also agree well with \cite{Sun2025ApJ...980...12S}, who obtained accurate spectroscopic redshifts and near-infrared photometry from JWST NIRCam grism spectroscopy to supplement their ALMA data. 

The agreement between our LFs and these ALMA works at higher redshifts again confirms the validity of using MIR observations as a reliable proxy for $L_{\rm IR}$ in star-forming galaxies, a point discussed in \citetalias{Ling2024} and Section \ref{sec:sed_fitting}. 
It is worth noting that, given their comparable depth at these redshifts, combining deep ALMA observations with JWST data holds significant promise for significantly reducing uncertainties in SED fitting and LF derivation in future studies, as demonstrated by \cite{Sun2025ApJ...980...12S}.

\subsection{AGN LFs}
\label{sec:AGN_LF}
In this section, we present the AGN LFs derived from the 19\% (534) sources identified as AGN hosts in our sample (Section \ref{sec:sed_fitting}).
The procedure for deriving the AGN LF is identical to that described for the $L_{\rm IR}$ LF (Section \ref{sec:compute_LF}), however, to simplify the subsequent calculation of the black hole accretion density (BHAD), we utilize the AGN luminosity ($L_{\rm AGN}$) rather than the IR luminosity ($L_{\rm IR}$).

\begin{figure}
    \centering
    \includegraphics[width=.5\textwidth] {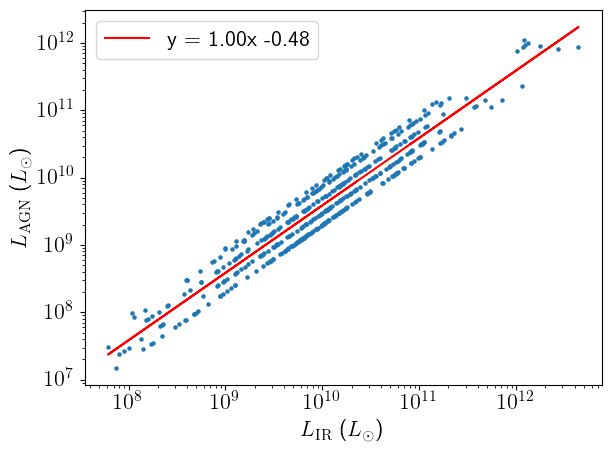}
    \includegraphics[width=.5\textwidth] {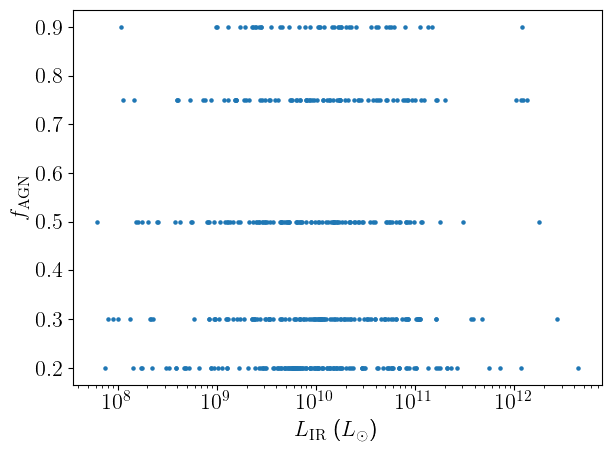}
    \caption{
    Upper: Relationship between the AGN luminosity ($L_{\text{AGN}}$) and the total infrared luminosity ($L_{\text{IR}}$) for our sample of AGN host galaxies, both derived from CIGALE. The red line represents the adopted best-fit conversion relation, $\log L_{\text{AGN}} = 1.00 \times \log L_{\text{IR}} - 0.48$.
    Lower: Scatter plot showing the total infrared luminosity ($L_{\text{IR}}$) versus the $\text{frac}_{\text{AGN}}$ derived from CIGALE of our AGN sample.
    }
    \label{fig:AGN_TIR}
\end{figure}

To allow direct and consistent comparison between our AGN luminosity functions and literature, we apply a standard conversion factor to estimate the AGN luminosity ($L_{\text{AGN}}$) from the total infrared luminosity ($L_{\text{IR}}$), $L_{\text{AGN}} = 10^{-0.48} \times L_{\text{IR}}$, or approximately $L_{\text{AGN}} = 0.33 \times L_{\text{IR}}$. 
This scaling factor originates from the average $\text{frac}_{\text{AGN}}$ measured for our AGN sample (Figure \ref{fig:AGN_TIR}, upper panel), where analysis shows $\text{frac}_{\text{AGN}}$ exhibits a uniform distribution in this sample and shows no significant correlation with redshift or $L_{\text{IR}}$ (Figure \ref{fig:AGN_TIR}, lower panel).

Although individual $L_{\text{AGN}}$ values are directly output by CIGALE, employing this average conversion factor ensures that $L_{\text{IR}}$ values from other AGN LF studies can be uniformly mapped to an equivalent $L_{\text{AGN}}$ scale.
Crucially, because the luminosity function is a statistical measure considering binned average luminosities, the adopted scaling relation provides a reasonable correspondence for the mean $L_{\text{AGN}}$ within each $L_{\text{IR}}$ bin, and introduces no systematic bias to the final LF shape.
The limiting $L_{\rm IR}$ luminosity applicable to the AGN sample (using the Sey2 template, see Section \ref{sec:compute_LF}) is also scaled by this factor to define the corresponding AGN luminosity limit.

\begin{figure*}
    \centering
    \includegraphics[width=\textwidth] {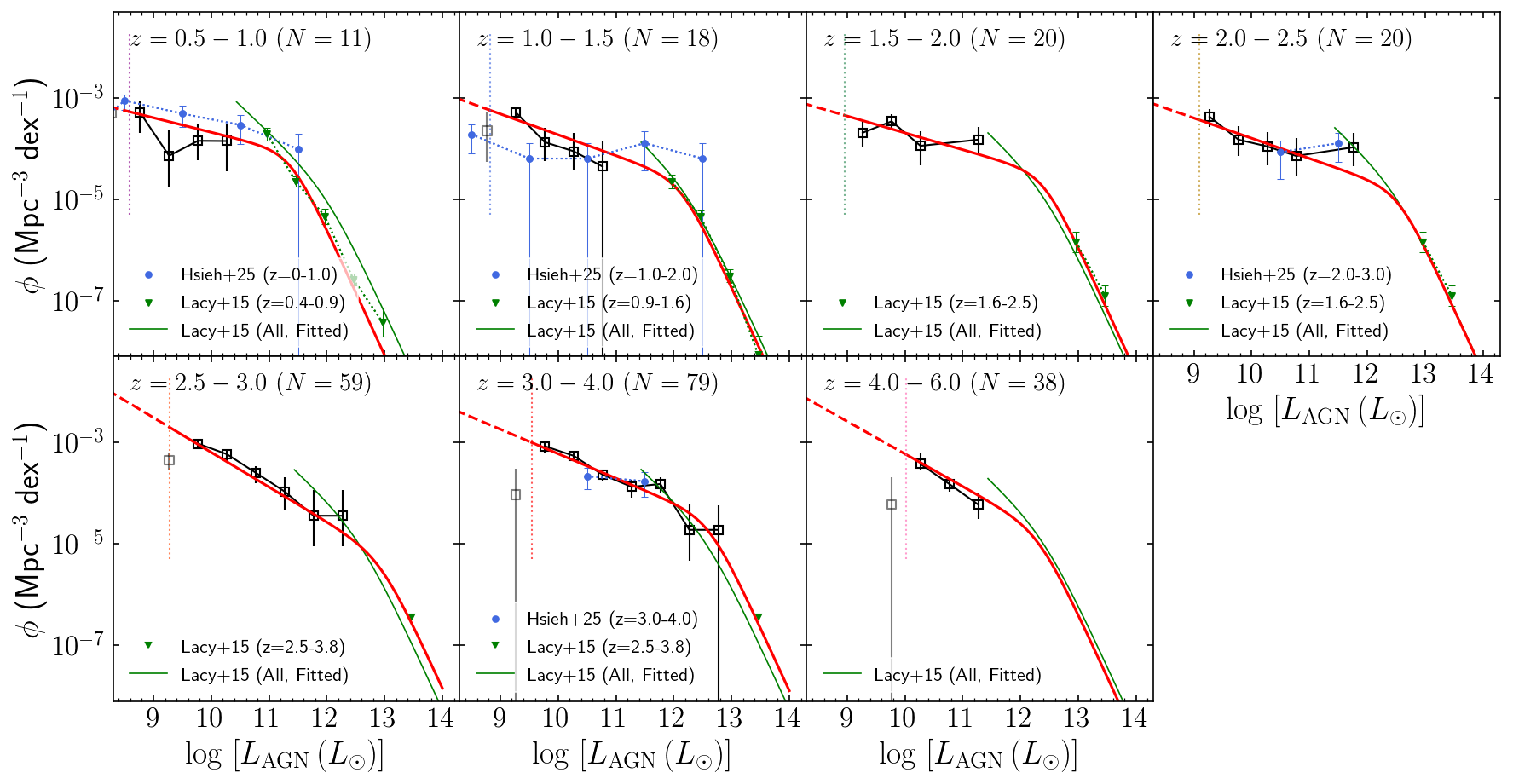}
    \caption{
    AGN luminosity functions (open black squares).
    Vertical dashed lines indicate the luminosity at which the completeness drops to 80\%. 
    The data points below the limits are shown in gray and are not connected. 
    The number of galaxies in the redshift bin is indicated at each panel's top. 
    For comparison, we show the AGN LFs for all populations from \citet[green triangles and green solid line]{Lacy2015ApJ...802..102L} and \citet[blue circles]{Hsieh2025}. 
    The green line shows the double power-law fit to the \cite{Lacy2015ApJ...802..102L}, and the red line is the best fit to our data combined with \cite{Lacy2015ApJ...802..102L}. 
    Fits below the luminosity limit are represented by dashed lines.
    }
    \label{fig:LF_AGN}
\end{figure*}

Figure \ref{fig:LF_AGN} presents our derived AGN LFs. 
For comparison, we include results from Spitzer \citep[green triangles;][]{Lacy2015ApJ...802..102L} and a recent JWST study \citep[blue circles;][]{Hsieh2025}. 
\cite{Lacy2015ApJ...802..102L} constructed a sample of 479 confirmed AGN selected at 24 $\mu$m, encompassing various populations (Type 1, Red Type 1, Type 2); here we plot their total (All) AGN LF. 
\cite{Hsieh2025} represents the first AGN LF study based on JWST CEERS data, using a sample of 41 AGN (selected from \citetalias{Ling2024} and \citealp{Chien2024MNRAS.532..719C}) detected in at least three MIRI bands. 

The luminosity range and evolutionary trend of our AGN LFs are similar to and consistent with \cite{Hsieh2025}, covering $L_{\rm AGN} \approx 10^{9} - 10^{10.5} L_\odot$ in the lowest redshift bin ($z=0.5-1.0$) and extending up to $L_{\rm AGN} \approx 10^{12} L_\odot$ at intermediate and higher redshifts ($z>2$). 
Benefiting from our significantly larger sample size ($\sim 13 \times$ larger), we achieve a substantial reduction in uncertainties compared to \cite{Hsieh2025}, allowing for a more refined determination of the LF shape within each redshift bin. 
This improved precision enables the first robust constraints on the faint-end slope of the AGN LF across multiple redshift bins, completing our understanding of the number density evolution of the low-luminosity AGN population.

It is worth mentioning that the CIGALE parameter space applied in both this work and \cite{Hsieh2025} (following \citetalias{Ling2024} and \citealp{Yang2023ApJ...950L...5Y}) favors selecting obscured (Type 2) AGN. 
This might lead to incompleteness regarding unobscured, face-on (Type 1) AGN and underestimation of the true AGN number density in our LF, particularly at the brighter end, where Type 1 AGN are more prevalent. 
However, \cite{Lacy2015ApJ...802..102L} has suggested that Type 2 AGN (along with a few partially obscured Type 1.X AGN) dominate the AGN population at lower luminosities ($L_{\rm AGN} < 10^{12} L_\odot$). 
As our study primarily probes this luminosity regime, this selection bias is not expected to impact our conclusions regarding the faint-end evolution significantly.
Figure \ref{fig:LF_AGN} also displays functional fits derived from \cite{Lacy2015ApJ...802..102L} alone, as well as a joint fit combining their data with ours. 
These fits will be discussed in Section \ref{sec:LF_fitting}. 

\subsection{Functional fitting}
\label{sec:LF_fitting}
We introduced two formalizations to fit and examine the evolution of LFs quantitatively.
For $L_{\rm IR}$ LF, we adopt the modified-Schechter function \citep{Saunders1990}:
\begin{align}\label{eq:modSfun}
    \phi(L)d \log L &= \phi^* \left( \frac{L}{L^*} \right)^{1-\alpha} \nonumber\\
    &\times \exp \left[ -\frac{1}{2\sigma^2} \log^2_{10} \left( 1+\frac{L}{L^*} \right) \right] d \log L
\end{align}
which exhibits a power-law behavior $\propto (L/L^*)^{1-\alpha}$ at the faint end ($L < L^*$), and a Gaussian roll-off at the bright end ($L > L^*$), with the Gaussian width parameterized by $\sigma$. The parameter $\phi^*$ represents the normalization, indicating the characteristic density of galaxies.

For AGN LF, we utilize a double power-law function, which is more appropriate \citep[e.g.][]{Lacy2015ApJ...802..102L} for describing the behavior and steep drop of the AGN number density:
\begin{align}
    \phi(L)d \log L &= \phi^* \left[\left(\frac{L}{L^*}\right)^{\gamma_1} + \left(\frac{L}{L^*}\right)^{\gamma_2}\right]^{-1} d \log L
    \label{eq:DPL}
\end{align}
where $\gamma_1$ and $\gamma_2$ determine the slope of the two sides of LF.

Our fitting procedure mainly follows \citetalias{Ling2024}, employing the Markov chain Monte Carlo (MCMC) method by the Python package emcee \citep{EMCEE2013PASP..125..306F}.
A key difference is that we incorporate data points from \cite{Gruppioni2013MNRAS.432...23G} (for the $L_{\rm IR}$ LF) and \cite{Lacy2015ApJ...802..102L} (for the AGN LF). 
This takes advantage of the large statistical samples from these studies to supplement our JWST data at the bright end, enabling a more reliable fit, particularly around the LF knee.
Notably, both our SMILES data and the \cite{Gruppioni2013MNRAS.432...23G} data target the GOODS-S field, which ensures consistency regarding cosmic variance when combining the datasets (see also the agreement in Figure \ref{fig:LF_TIR}).

The specific steps of our fitting procedure are outlined as follows:
First, data points from the literature LF are merged into our LF bins based on their respective redshift ranges. 
For luminosity bins where the datasets overlap, we decided the error bars to be the full range covered by the minimum and maximum uncertainties from both datasets.
As shown in Figures \ref{fig:LF_TIR} and \ref{fig:LF_AGN}, data points below the limiting luminosity (shown in gray) are excluded from the fit.
The bright-end slopes for both LFs are directly fixed to the values found in the literature: $\sigma=0.5$ for the $L_{\rm IR}$ LF \citep{Gruppioni2013MNRAS.432...23G} and $\gamma_2 = 2.48$ for the AGN LF \citep{Lacy2015ApJ...802..102L}, as these parameters are already well-constrained by those studies and are not the primary focus here.

\begin{figure}
    \centering
    \includegraphics[width=.5\textwidth] {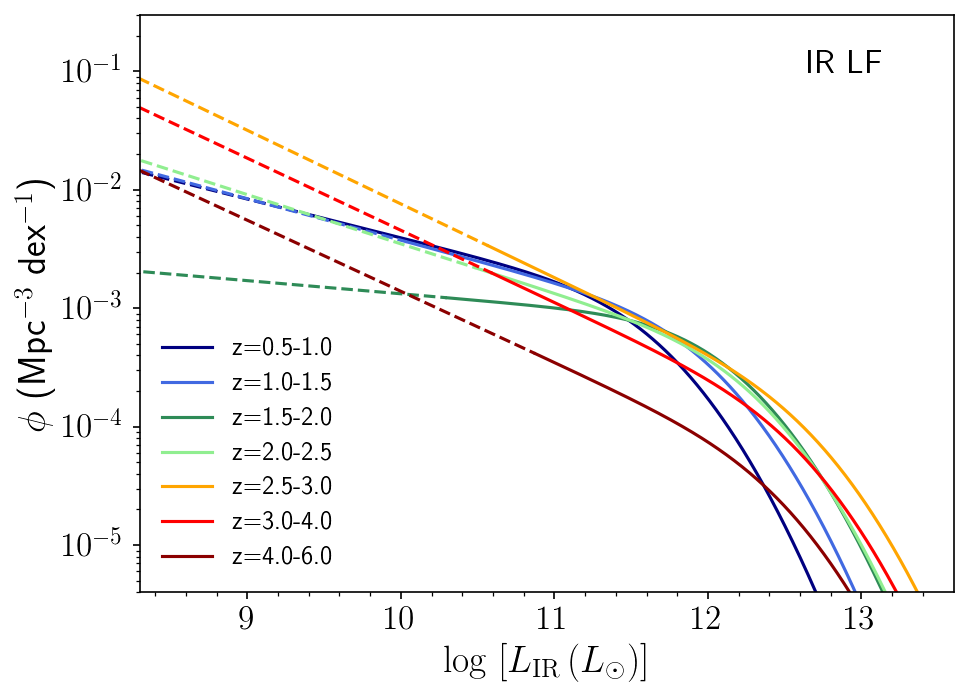}
    \includegraphics[width=.5\textwidth] {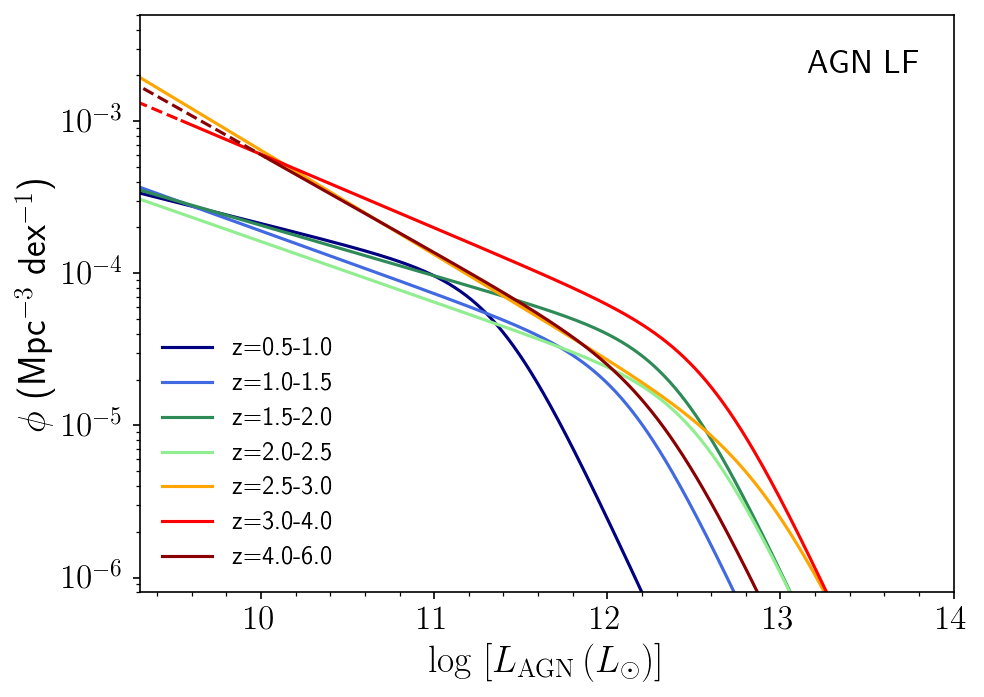}
    \caption{Best-fit luminosity functions for infrared ($L_{\rm IR}$, upper panel) and AGN (lower panel) emission.
    The solid lines represent the best-fit (median) curves derived from the MCMC fitting procedure. Curves below the luminosity limit are represented by dashed lines.}
    \label{fig:LF_ALL_fit}
\end{figure}

We then perform the MCMC fitting for each LF, fitting the remaining three free parameters in Equations \ref{eq:modSfun} and \ref{eq:DPL}.
We configured the MCMC with 100 walkers and ran the sampler for 10,000 steps. The initial 1,000 steps are discarded as burn-in to ensure convergence.
For the $L_{\rm IR}$ LF, flat priors are set with ranges of $\log (L^*)=[8, 13]$, $\log (\phi^*)=[-5, -1]$, and $\alpha=[-1,3]$.
For the AGN LF, we set the flat prior ranges as $\log (L^*)=[10, 14]$, $\log (\phi^*)=[-7, -3]$, and $\gamma_1=[0.1,1.0]$.
The best-fit parameters (median) and their uncertainties (16th and 84th percentiles) for each LF are provided in Table \ref{tab:lf_fits_combined}.
Figure \ref{fig:LF_ALL_fit} presents the best-fit curves for the $L_{\rm IR}$ LF and AGN LF, respectively, across the different redshift bins. 

\begin{deluxetable}{ccccc}
\tablewidth{0pt}
\tablecaption{Best-fit Parameters for the infrared and AGN Luminosity Functions \label{tab:lf_fits_combined}}
\tablehead{
    \colhead{\textbf{Redshift Bin}} & & & & \\
    \textbf{$L_{\rm IR}$ LF (Modified Schechter)} &
    \colhead{$\log (L^*/L_\odot)$} & 
    \colhead{$\log (\phi^*/{\rm Mpc^{-3}\,dex^{-1}})$} & 
    \colhead{$\alpha$} & 
    \colhead{$\sigma$ (fixed)}
}
\startdata
$0.5 < z < 1.0$ & $11.16^{+0.16}_{-0.14}$ & $-2.78^{+0.17}_{-0.20}$ & $1.33^{+0.12}_{-0.13}$ & $0.5$ \\
$1.0 < z < 1.5$ & $11.48^{+0.13}_{-0.14}$ & $-2.95^{+0.15}_{-0.16}$ & $1.35^{+0.10}_{-0.12}$ & $0.5$ \\
$1.5 < z < 2.0$ & $11.57^{+0.32}_{-0.32}$ & $-3.05^{+0.18}_{-0.31}$ & $1.11^{+0.30}_{-0.38}$\tablenotemark{a} & $0.5$ \\
$2.0 < z < 2.5$ & $11.80^{+0.20}_{-0.22}$ & $-3.21^{+0.22}_{-0.23}$ & $1.42^{+0.16}_{-0.21}$ & $0.5$ \\
$2.5 < z < 3.0$ & $12.23^{+0.24}_{-0.21}$ & $-3.50^{+0.31}_{-0.37}$ & $1.62^{+0.15}_{-0.17}$ & $0.5$ \\
$3.0 < z < 4.0$ & $12.14^{+0.28}_{-0.25}$ & $-3.65^{+0.26}_{-0.30}$ & $1.62^{+0.12}_{-0.13}$ & $0.5$ \\
$4.0 < z < 6.0$ & $12.02^{+0.67}_{-0.63}$ & $-4.07^{+0.44}_{-0.41}$ & $1.6$\tablenotemark{b} & $0.5$ \\
\hline
\textbf{AGN LF (Double Power-Law)} & & & $\gamma_1$ & $\gamma_2$ (fixed) \\
\hline
$0.5 < z < 1.0$ & $11.38^{+0.14}_{-0.13}$ & $-4.07^{+0.24}_{-0.25}$ & $0.29^{+0.18}_{-0.12}$ & $2.48$\\
$1.0 < z < 1.5$ & $12.14^{+0.17}_{-0.17}$ & $-4.60^{+0.35}_{-0.32}$ & $0.41^{+0.12}_{-0.13}$ & $2.48$\\
$1.5 < z < 2.0$ & $12.41^{+0.34}_{-0.21}$ & $-4.48^{+0.44}_{-0.72}$ & $0.33^{+0.22}_{-0.16}$ & $2.48$\\
$2.0 < z < 2.5$ & $12.55^{+0.38}_{-0.25}$ & $-4.81^{+0.55}_{-0.75}$ & $0.40^{+0.20}_{-0.18}$ & $2.48$\\
$2.5 < z < 3.0$ & $12.93^{+0.54}_{-0.37}$ & $-5.20^{+0.77}_{-0.88}$ & $0.68^{+0.16}_{-0.21}$ & $2.48$\\
$3.0 < z < 4.0$ & $12.63^{+0.25}_{-0.22}$ & $-4.48^{+0.35}_{-0.42}$ & $0.48^{+0.14}_{-0.13}$ & $2.48$\\
$4.0 < z < 6.0$ & $12.33^{+1.13}_{-1.08}$ & $-4.71^{+0.74}_{-1.00}$ & $0.64^{+0.24}_{-0.31}$ & $2.48$\\
\enddata
\tablenotetext{a}{$\alpha=1.62^{+0.09}_{-0.10}$ if include the bin at the luminosity limit.}
\tablenotetext{b}{For the $L_{\rm IR}$ LF in the highest redshift bin ($4.0 < z < 6.0$), $\alpha$ could not be constrained reliably due to limited data points at the faint end and was fixed to $\alpha=1.6$ assuming the same faint end slope as previous redshift bin.}
\end{deluxetable}

Simultaneously fitting $\phi^*$, $L^*$, and the faint-end slope ($\alpha$ or $\gamma_1$) is unique to this study. 
Previous IR LF works, such as \cite{Goto2010A&A...514A...6G} or \cite{Gruppioni2013MNRAS.432...23G}, were often limited by their data depth, forcing the assumption of a fixed faint-end slope which precluded investigation into its potential evolution.
Benefiting from the large sample size provided by SMILES, we are able, for the first time, to observe the evolution of the faint-end slopes of LF with redshift.

For the $L_{\rm IR}$ LF, $\alpha$ is found to be steeper at $z > 2.5$ ($\sim 1.6$), flattens slightly starting from cosmic noon ($z < 2.5$, $\alpha \sim 1.3-1.4$), and remains the flatter slope all through the lowest redshift bin ($z=0.5-1.0$).
We remark that $\alpha$ deviates from the trend, dropping to 1.1 at $z=1.5-2.0$. This is because the faint-end data point, which lies just below our limiting luminosity, was excluded during the fitting of this redshift bin. Should the point be included, the derived slope would be $\alpha=1.62^{+0.09}_{-0.10}$, more consistent with higher redshifts.

While the difference in $\alpha$ is only statistically significant around $z \approx 2.5$ (Table \ref{tab:lf_fits_combined}), this evolution may indicate different phases of the Universe in the efficiency of forming new galaxies, suggesting that the formation of new, low-luminosity galaxies has significantly slowed down after cosmic noon ($z<2$).
On the other hand, such faint galaxies formed at higher redshifts may have evolved, potentially through mergers, moving towards the LF knee (around $L^* \approx 10^{11} L_\odot$) by $z=0.5-1.0$, contributing to its prominence.

Although we report a flatter faint-end slope $\alpha$, especially at low redshifts, our results are largely compatible with past LFs. 
This is because their luminosity range typically either does not extend beyond the knee of the LF \citep[e.g.,][]{Gruppioni2013MNRAS.432...23G, Magnelli2013A&A...553A.132M} or exhibits substantial fitting uncertainties (e.g., \citetalias{Ling2024}), and simply fix $\alpha$ at $\sim1.5$.
It is worth noting that this value is close to the average value we find across all redshifts; therefore, such variations could be averaged out.
Direct comparison with our LF and JWST data points from \citetalias{Ling2024} suggests that they are actually similar in terms of faint-end slope.

We do observe a distinct difference between our faint end and that of \cite{Fujimoto2024ApJS..275...36F}. \cite{Fujimoto2024ApJS..275...36F} derives LF from a blind sample of extremely faint 1.2 mm sources with lensing magnification, and achieves a comparable luminosity limit to ours at $L_{\rm IR} \approx 10^{10} L_\odot$.
They reported a much steeper and constant slope of $\alpha\sim2$ (compared to ours $\sim1.3-1.6$), which also disagrees with past IR LFs \citep[e.g.,][]{Gruppioni2013MNRAS.432...23G, Gruppioni2020A&A...643A...8G} and simulations \citep[e.g.,][]{Lagos2020MNRAS.499.1948L}. 
They show that this discrepancy remains remarkable after accounting for systematics such as cosmic variance, lensing models, and the sample bias.

Nevertheless, we find that their LF jumps and exhibits uncertainties of up to 0.5 dex at the luminosity limit, allowing for a considerable range of $\alpha$.
By comparing each redshift bin individually, their $\alpha$ at $z=0.6-1.0$ are $1.94\pm0.39$ (ours: $1.33+0.12$), $1.94\pm0.43$ at $z=1.0-2.0$ (ours: $1.35+0.10$ and $1.62+0.09$), $1.93\pm0.26$ at $z=2.0-3.0$ (ours: $1.42+0.16$ and $1.62+0.15$), and $2.04\pm0.20$ (ours: $1.62+0.12$) for $z=3.0-4.0$. 
Except for the lowest redshift, the lower/upper limits of the two are in fact nearly overlapping.

Greater systematic errors may come from the different physical mechanisms traced by MIR and sub-mm observations.
ALMA 1.2mm primarily observes the tail of dust thermal radiation (i.e., cold dust), while MIRI detects hotter dust heated by young stars, also capturing AGN dusty torus.
For brighter or more massive galaxies, the connection between luminosity at these wavelengths ($L_{\rm MIR}$ and $L_{\rm FIR}$) and total IR luminosity $L_{\rm IR}$ has been well-modeled, so we do not observe significant discrepancies (e.g., our results match all three ALMA LFs at the bright end, including \citealt{Fujimoto2024ApJS..275...36F}).
For low-mass galaxies, however, our current understanding of the dependence between dust and metallicity may still be incomplete \citep{Fujimoto2024ApJS..275...36F}, and dusty star formation may also exhibit different behaviors across such a wide redshift range \citep[e.g.,][]{Casey2014PhR...541...45C}.
Further integration of JWST+ALMA joint deep observations will help elucidate how dust evolution drives these divergences.

In terms of overall number density, the LF peaks at $z=2.5-3.0$ (orange line in the upper panel of Figure \ref{fig:LF_ALL_fit}), coinciding with the known peak of cosmic star formation history.
Despite the clear increase in the number density of IR galaxies from $z=6.0$ to $z=2.5$, the evolution of the normalization parameters $L^*$ and $\phi^*$ appears less distinctive. 
Both parameters change by only $\sim 1$ dex across the entire redshift range studied, and show relatively large uncertainties.
As discussed in \citetalias{Ling2024}, a strong degeneracy between $L^*$ and $\phi^*$ is expected and observed through the MCMC analysis. 
Even with the inclusion of literature data (to better constrain the bright end), significant uncertainties remain around the LF knee ($L_{\rm IR} \sim 10^{11}-10^{12} L_\odot$), making it challenging to determine the evolution of $L^*$ and $\phi^*$ individually.
It is worth noting, however, that this degeneracy would not affect the calculation of the luminosity density since the overall shape of the LF has been well-determined by the combined dataset.

We also observe a similar evolution in the best-fit faint-end slope $\gamma_1$ for the AGN LF.
As demonstrated in Figure \ref{fig:LF_AGN}, the faint-end slope from \cite{Lacy2015ApJ...802..102L} appears to be poorly constrained (and fixed), relying on only one data point or the extrapolations from the bright end.
Based on our low-luminosity AGN sample, we present a more robust $\gamma_1$ that is smaller than their measurement ($\sim 1$). The values span from 0.6 to 0.3 among all redshift ranges.
The decrease of $\gamma_1$ happens around $z=2.5$, although it is not statistically significant given the current uncertainties. 
At $z < 2.5$, $\gamma_1$ fluctuates around $\sim 0.3-0.4$, without an apparent flattening seen in the $L_{\rm IR}$ LF at low redshifts.
The number density evolution at the faint end suggests that faint and/or highly obscured AGN might have been active from high redshifts $z \sim 6$ down to $z = 2.5$.

The other evolution observed in our AGN LF is the decrease in $L^*$ with decreasing redshift, a trend already reported by \cite{Lacy2015ApJ...802..102L} (we note their fitting methodology differed, fixing $\phi^*$ and the slopes while allowing $L^*$ to evolve with a cubic expression, following \citealt{Hopkins2007ApJ...654..731H}).
As with the $L_{\rm IR}$ LF, the evolution of $L^*$ and $\phi^*$ for the AGN LF appears minor (within $\sim 1$ dex) and exhibits significant degeneracy and fluctuations across redshift bins, making it difficult to discern a clear evolutionary pattern beyond the established trend in $L^*$.

\section{Luminosity density}
\label{sec:ld}
By integrating our best-fit LFs $\phi(L)$ (Equations \ref{eq:modSfun} and \ref{eq:DPL}) over luminosity, we derive the luminosity density (LD; $\rho_L$):
\begin{align}
    \rho_L &= \int^{10^{15}}_{10^8} \phi (L) \, L \, d \log L
\end{align}
which represents the energy density emitted by galaxies per unit comoving volume at a given redshift, corresponding to the infrared ($L_{\rm IR}$) and AGN luminosity in our analysis. 
While our data are complete to a luminosity limit of $L_*\sim10^9$ $L_\odot$, we adopt conservative integration limits down to $10^8$, including the extrapolated contribution from the faint-end of the galaxy population. This allows us to compare with literature \citep[e.g.][]{Gruppioni2013MNRAS.432...23G, Gruppioni2020A&A...643A...8G} that adopts this convention directly.
To examine the LD uncertainty that may be introduced by extrapolation, we also performed a separate integration directly on the available part of the LF. 
Specifically, we integrate the sensitive luminosity range above the completeness limit within each LF (roughly from $10^9$ to $10^{12}$ $L_\odot$). 

Following \citetalias{Ling2024}, the value of $\rho_L$ and its $1\sigma$ uncertainty are determined from the median, 16th, and 84th percentiles of the probability distribution obtained by integrating each MCMC model fit.
We compute $\rho_{L_{\rm IR}}$ and $\rho_{L_{\rm AGN}}$ for the $L_{\rm IR}$ LF and AGN LF (Figure \ref{fig:LF_ALL_fit}) in each redshift bin. 
Table \ref{tab:ld} lists the derived luminosity densities.
Thanks to the well-fit LF, the uncertainty of derived LD is considerably small among almost all the redshift ranges.

\begin{deluxetable}{ccc}
\tablewidth{0pt}
\tablecaption{Luminosity densities (LD), derived from an extrapolation of LF at $L_{\rm IR} = 10^8 - 10^{15}$ $L_\odot$}\label{tab:ld}
\tablehead{
\textbf{Redshift Bin} & $\log (\rho_{L_{\rm IR}} / (L_\odot {\rm Mpc}^{-3}))$ & $\log (\rho_{L_{\rm AGN}} / (L_\odot {\rm Mpc}^{-3}))$
}
\startdata
$0.5 < z < 1.0$ & $8.59^{+0.06}_{-0.06}$ & $7.19^{+0.10}_{-0.10}$ \\
$1.0 < z < 1.5$ & $8.76^{+0.04}_{-0.04}$ & $7.47^{+0.14}_{-0.12}$ \\
$1.5 < z < 2.0$ & $8.81^{+0.09}_{-0.07}$ & $7.81^{+0.21}_{-0.26}$ \\
$2.0 < z < 2.5$ & $8.85^{+0.06}_{-0.05}$ & $7.66^{+0.24}_{-0.26}$ \\
$2.5 < z < 3.0$ & $8.99^{+0.06}_{-0.05}$ & $7.91^{+0.23}_{-0.19}$ \\
$3.0 < z < 4.0$ & $8.76^{+0.09}_{-0.08}$ & $8.11^{+0.15}_{-0.14}$ \\
$4.0 < z < 6.0$ & $8.20^{+0.26}_{-0.22}$ & $7.58^{+0.48}_{-0.31}$ \\
\enddata
\end{deluxetable}

\begin{figure*}
    \centering
    \includegraphics[width=\textwidth] {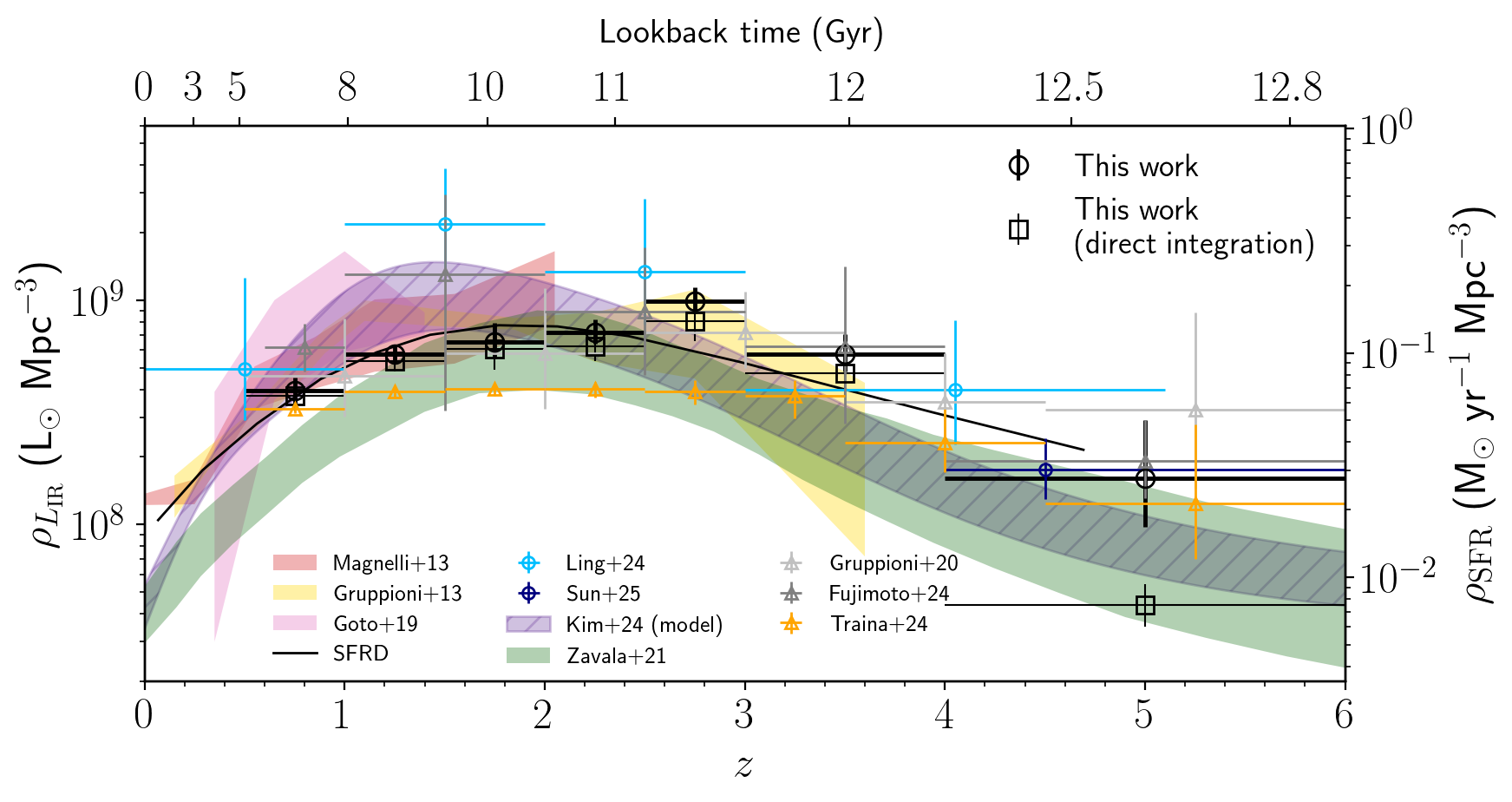}
    \caption{
        Redshift evolution of infrared luminosity density, $\rho_{L_{\rm IR}}$ (open circle). 
        Thin squares represent $\rho_{L_{\rm IR}}$ derived from the direct integration of LF.
        The horizontal error bars indicate the width of redshift bins. 
        $\rho_{L_{\rm IR}}$ evolution from the literature \protect \citep{Magnelli2013A&A...553A.132M,  Gruppioni2013MNRAS.432...23G, Goto2019PASJ...71...30G, Gruppioni2020A&A...643A...8G, Zavala2021ApJ...909..165Z, Kim2024MNRAS.527.5525K, Fujimoto2024ApJS..275...36F, Traina2024A&A...681A.118T, Sun2025ApJ...980...12S} are also provided. 
        Note that we take the value that integrated down to 10$^8$ L$_\odot$ for \citet{Fujimoto2024ApJS..275...36F} for fair comparison.
        The black line represents the star formation rate density (SFRD) from \citet{Madau2014ARA&A..52..415M}.
        SFR density ($\rho_{\rm SFR}$) converted from $\rho_{\rm L_{IR}}$ is shown for reference on the right, assuming the relation from \protect\cite{Kennicutt1998}. }
    \label{fig:A}
\end{figure*}

Figure~\ref{fig:A} presents the redshift evolution of $\rho_{L_{\rm IR}}$, i.e., dusty cosmic star formation rate density (CSFRD). 
For reference, we also show SFR density ($\rho_{\rm SFR}$) at the right. $\rho_{\rm SFR}$ is converted from $\rho_{\rm L_{IR}}$ assuming relation from \cite{Kennicutt1998}:
\begin{align}
    \rm SFR \, [M_\odot yr^{-1}] = 1.72 \times 10^{-10} \,L_{\rm IR}^{\rm tot}\,[L_\odot]
\end{align}
Our measurements, shown as black circles, align remarkably well with the star formation rate density from \cite{Madau2014ARA&A..52..415M} and ALMA-based results (shown by triangles) from \citet{Gruppioni2020A&A...643A...8G}, \citet[while their result is slightly lower than all the other IR measurements at $z<1.5$]{Zavala2021ApJ...909..165Z}, \citet{Fujimoto2024ApJS..275...36F}, \citet{Traina2024A&A...681A.118T}, and \citet[navy circles]{Sun2025ApJ...980...12S}, especially at $z>2.5$.
At $z=5$, our measurement is close to and has the same uncertainty level as \citet{Sun2025ApJ...980...12S}, which combines JWST NIRSpec and ALMA photometry.
This agreement likely reflects the sensitivity of both methods to obscured star formation, and again, demonstrates the capability of MIR photometry to independently derive IR luminosity without FIR observations.
In addition, we compare our LD (which takes the extrapolation of the LF) to the direct integration results (thin black squares).
We notice that the LD based on the direct integration is only slightly lower than that from our extrapolation method. 
A significant gap appears only at the highest redshifts ($z=4-6$), due to the insufficient number of available data points in the LF.

We notice both \citet{Gruppioni2013MNRAS.432...23G, Gruppioni2020A&A...643A...8G} and our study find a CSFRD peak at $z\sim 2.5-3$, slightly higher than most optical/UV measurements, though their error bars are substantial. 
The flattened shape of CSFRD at this redshift range, as suggested by \citet{Traina2024A&A...681A.118T}, could reflect the formation of massive spheroids of elliptical galaxies at an earlier cosmic age \citep{Calura2003ApJ...596..734C}.
However, interestingly, a discrepancy arises between our measurements at $z=1-2$ and those previously done in the CEERS field (\citetalias{Ling2024}), where our data points fall below at their lower end.
We speculate that this difference stems from cosmic variance, as the same analysis method is employed in both works, and the discrepancy was already apparent in the MIRI number counts between these two fields.
\citet{Moster2011ApJ...731..113M} has shown that down to 10$^{10}$ $M_\odot$, the relative cosmic variance of GOODS field at $z=1-2$ is $15-20$\%. 
As the value was estimated with an area of 320 arcmin$^2$ ($\sim 10$ times larger than SMILES), the actual variance should be larger, and typical clusters and filaments with a size of several Mpc can easily affect our observed field. 
Observations in yet another field with MIRI filter coverage would help resolve this issue.

\begin{figure*}
    \centering
    \includegraphics[width=\textwidth] {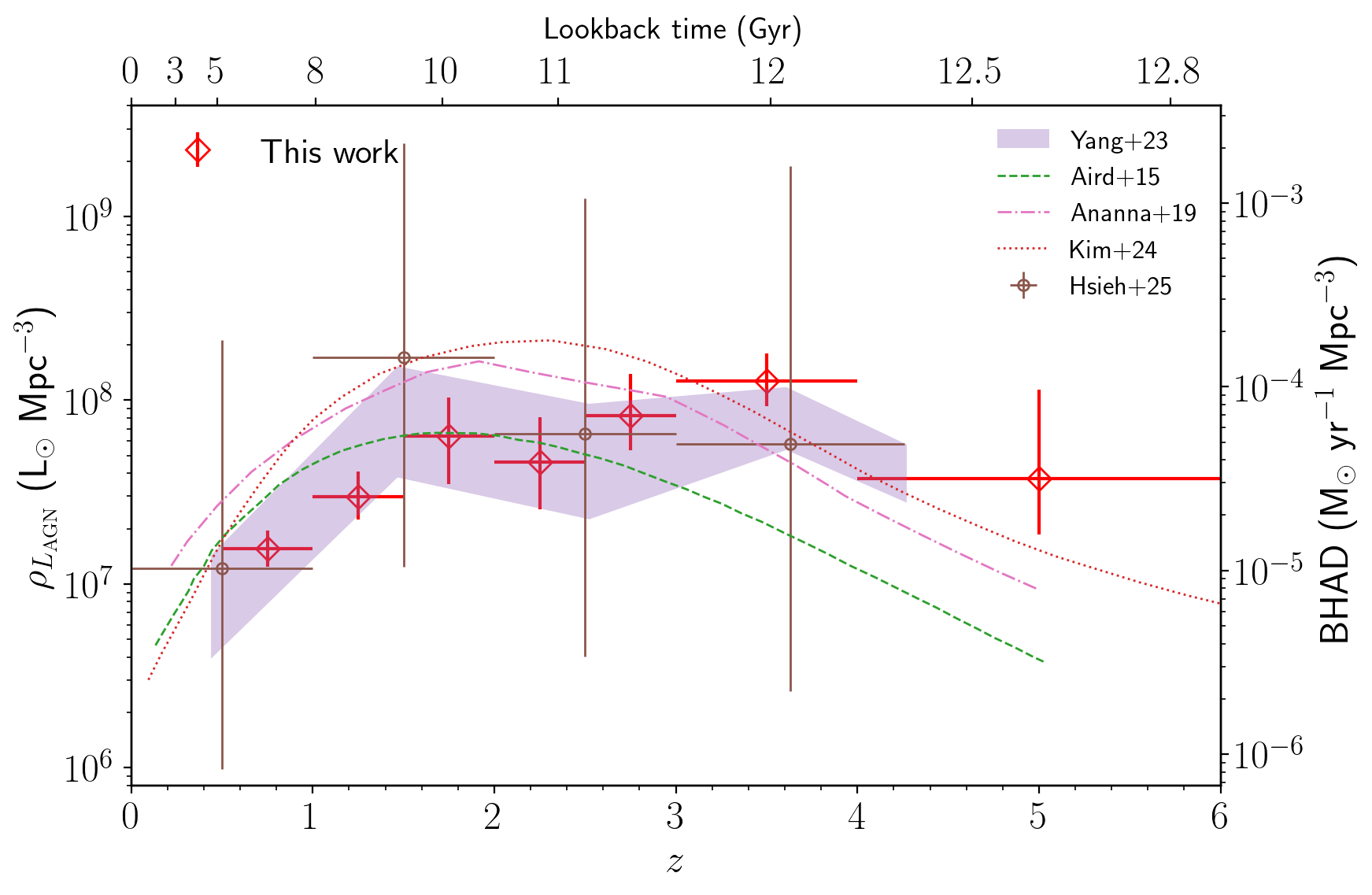}
    \caption{
        Redshift evolution of BHAD (red open diamond). 
        The horizontal error bars indicate the width of redshift bins. 
        The dashed, dash-dotted, and dotted lines denote the two X-ray and one MIR BHAD as reported in previous studies ~\citep{Aird2015MNRAS.451.1892A, Ananna2019ApJ...871..240A, Kim2024MNRAS.527.5525K}, respectively. 
        The small data points (brown open circles) are based on the JWST-selected AGNs in the CEERs field \citep{Hsieh2025}.
        }
    \label{fig:BHAD}
\end{figure*}

In Figure \ref{fig:BHAD}, we present new measurements of the AGN luminosity density ($\rho_{L_{\rm AGN}}$) and the corresponding black hole accretion density (BHAD) evolution.
We briefly introduce the literature results used for comparison:
\cite{Aird2015MNRAS.451.1892A} constructed AGN X-ray LFs from Chandra soft/hard X-ray data and derived the BHAD out to $z \sim 5$.
\cite{Ananna2019ApJ...871..240A} employed a new population synthesis model to estimate the total X-ray emission from SMBH growth, reproducing observational constraints from various X-ray surveys on the unresolved X-ray background (XRB), number counts, and X-ray LFs, thereby estimating the BHAD.
\cite{Kim2024MNRAS.527.5525K} considered multiple galaxy populations, used the evolutionary functions from \cite{Gruppioni2011MNRAS.416...70G}, and applied them to JWST mid-infrared number count constraints (\citealp{Ling2022MNRAS.517..853L}, \citealp{Wu2023MNRAS.523.5187W}), parametrically deriving the AGN accretion history.
\cite{Yang2023ApJ...950L...5Y}, also based on JWST MIR observations, directly summed the disk luminosities ($L_{\rm Disk}$) of AGN in the CEERS field to provide a conservative lower limit on the BHAD.
\cite{Hsieh2025} also utilized the CEERS field, but, as mentioned earlier, first calculated the AGN LF and then derived the BHAD, following a method identical to this work.
The figure also shows the cosmic star formation history (CSFH) from \cite{Madau2014ARA&A..52..415M}, scaled for comparison.

The conversion from $\rho_{L_{\rm AGN}}$ to BHAD follows the procedure used by \cite{Yang2023ApJ...950L...5Y} and \cite{Hsieh2025}. First, the black hole accretion rate (BHAR) is calculated:
\begin{align}
    \mathrm{BHAR} &= \frac{L_{\rm AGN} (1-\epsilon)}{\epsilon c^2}
\end{align}
The BHAD is then obtained by integrating the BHAR weighted by the AGN LF:
\begin{align}
    \mathrm{BHAD} &= \int \phi (L_{\rm AGN}) \, \mathrm{BHAR} \, d \log L_{\rm AGN} \\
        &= \int \phi (L_{\rm AGN}) \, \frac{L_{\rm AGN} (1-\epsilon)}{\epsilon c^2} \, d \log L_{\rm AGN} \nonumber\\
        &= \frac{1-\epsilon}{\epsilon c^2} \rho_{L_{\rm AGN}}
\end{align}
where $\epsilon$ is the radiative efficiency, assumed to be 0.1 (following \citealp{Brandt2015A&ARv..23....1B}), and $c$ is the speed of light.

We find that our results at $z<3$ are in excellent agreement with JWST works, i.e., \citet[]{Yang2023ApJ...950L...5Y} and \citet[]{Hsieh2025}.
The consistency with the preliminary data from \citet[]{Hsieh2025} provides a valuable cross-validation of our measurement.

One notable feature of our results is the detailed structure resolved around the peak of cosmic activity (cosmic noon). 
Rather than a single, broad peak, our data confirms the trend independently identified by \citet[]{Yang2023ApJ...950L...5Y}: a moderate evolution at $z=1-5$. 
Such a flattened curve may hint at and reflect a gradual transition in the dominant modes of AGN fueling. 
For instance, different populations of AGNs, driven by major mergers versus secular disk instabilities, could peak in activity at slightly different cosmic epochs \citep[e.g.,][]{Hopkins2009ApJ...694..599H, Ueda2014ApJ...786..104U, Aird2015MNRAS.451.1892A}.

Another implication of our BHAD evolution is the constraints placed on the early universe ($z > 3$). 
Our data point at $z = 5$ indicates a substantially higher BHAD than expected by X-ray works \citep{Aird2015MNRAS.451.1892A, Ananna2019ApJ...871..240A}, which feature a steep decline after $z\sim 3$, but align better with JWST IR models from \citet{Kim2024MNRAS.527.5525K} predicting more high-redshift activity.
This is mainly because MIR-selected AGN, in general, are highly obscured in the X-ray. 
As discussed by \citet{Yang2023ApJ...950L...5Y}, the obscuration especially occurs at high-$z$ where the abundant gas with high column density can easily damp the X-ray emission.
The missed AGN contribution at high-$z$ indicates that BHAD should already be considerable at $z\sim 5$, which may favor scenarios with more massive heavy BH seeds or simply, a more efficient early accretion - as already noticed by forefront JWST results. 
Future modeling and cosmological simulations should aim to reconcile these findings by revisiting the early BH growth pathways.

\section{Summary}
\label{sec:summary}
In this work, we have presented a comprehensive study of the infrared luminosity functions (LFs) of galaxies and active galactic nuclei (AGN) from $z=0.5$ to $z=6.0$. 
We utilized data from the SMILES survey, the most complete JWST MIRI imaging program to date, in conjunction with deep ancillary optical and near-IR photometry from the JADES program in the GOODS-S field. 
By the detailed SED fitting on a catalog of 2,607 reliable sources, we derived monochromatic LFs in eight MIRI bands, as well as infrared ($L_{\rm IR}$) and AGN LFs. 
We fit these LFs to quantify their evolution and subsequently derived the cosmic star formation rate density (CSFRD) and black hole accretion density (BHAD).

Our main findings are summarized as follows:
\begin{itemize}
    \item \textbf{Comprehensive Monochromatic LFs:} We have derived monochromatic LFs from 5.6 to 25.5 $\mu$m. Benefiting from the larger area and sample size of SMILES, our LFs have significantly smaller uncertainties and finer redshift binning than previous JWST studies. This work presents the first JWST-based LFs at 5.6 $\mu$m and 25.5 $\mu$m and pushes the constraints at 18, 21, and 25.5 $\mu$m to $z>3$ for the first time.
    
    \item \textbf{The First Robust Faint-End $L_{\rm IR}$ and AGN LF:} Our $L_{\rm IR}$ LFs extend to a limiting luminosity of $L_{\rm IR} \sim 10^{9} L_\odot$ at $z \sim 0.5-1.0$, providing some of the deepest constraints on the IR LF to date, especially at the faint end.
    Based on a sample of 534 AGN hosts, an order-of-magnitude larger than previous JWST studies, we are able to probe the first robust constraints on the faint-end slope ($\gamma_1$) of the AGN LF across multiple redshift bins, revealing it to be relatively flat ($\gamma_1 \approx 0.3-0.6$).    
    
    \item \textbf{Evolution of the Faint-End Slope:} Unlike previous studies that often fixed the faint-end slope, our data quality allowed us to constrain its evolution. For the $L_{\rm IR}$ LF, the faint-end slope ($\alpha$) steepens from $\alpha \approx 1.3$ below $z<2$ to $\alpha \approx 1.6$ after $z \sim 2.5-3.0$, suggesting that the formation of low-luminosity dusty galaxies was more efficient before cosmic noon than in the recent universe.
    
    \item \textbf{Cosmic Star Formation and Black Hole Growth:} Our derived CSFRD is in excellent agreement with results from ALMA and latest JWST works, with a flattened trend and peaking at $z\sim2.5-3.0$. 
    The BHAD derived from our well-constrained AGN LF may hint a complex evolution around Cosmic Noon and a slower decline at $z>3$ than found in previous X-ray-based studies, providing a crucial constraint for theoretical models to explain the rapid assembly of the first supermassive black holes.

\end{itemize}

The combination of wide-area MIRI surveys with deep multi-wavelength data has proven to be an exceptionally powerful tool for understanding the evolution of the dusty universe. 
Future progress will be driven by applying these methods to additional fields to mitigate cosmic variance and confirm the evolutionary trends reported here. 
The synergy with deep ALMA observations and extensive spectroscopic follow-up campaigns will be crucial for further refining SEDs and painting a complete picture of galaxy and black hole co-evolution through cosmic time.

\begin{acknowledgments}
The authors are grateful to the anonymous referee for the valuable comments, which significantly improved the paper. 
This work is based on observations made with the NASA/E-SA/CSA JWST. The data were obtained from the Mikulski Archive for Space Telescopes at the Space Telescope Science Institute, operated by the Association of Universities for Research in Astronomy, Inc., under NASA contract NAS 5-03127 for JWST. 
This work utilized high-performance computing facilities operated by the Center for Informatics and Computation in Astronomy (CICA) at National Tsing Hua University, funded by the Ministry of Education of Taiwan, the National Science and Technology Council of Taiwan, and the National Tsing Hua University.
TG acknowledges the support of the National Science and Technology Council of Taiwan through grants 113-2112-M-007-006, 113-2927-I-007-501, and 113-2123-M-001-008.
TH acknowledges the support of the National Science and Technology Council of Taiwan through grants 113-2112-M-005-009-MY3, 113-2123-M-001-008-, and 111-2112-M-005-018-MY3 and the Ministry of Education of Taiwan through the grant 113RD109. 
\end{acknowledgments}

\begin{contribution}

CTL led the project, performed the analysis, and wrote the manuscript. 
TG came up with the initial research concept and supervised the whole process. 
SJK assisted with the analysis procedure and LF calculation.
CW, AC, and TLP provided essential discussion and edited the figures.
EK helped obtain and process data.
TH validated the result.
All the co-authors assisted with the interpretation of results, discussion, and manuscript review. 
All authors approved the final version of the manuscript.


\end{contribution}

%
\facilities{HST, JWST}

\software{astropy \citep{astropy:2013, astropy:2018, astropy:2022},  
          CIGALE \citep{Boquien2019A&A...622A.103B}, 
          EAZY \citep{Brammer2008ApJ...686.1503B}
          }




\bibliography{PASPsample701}{}
\bibliographystyle{aasjournalv7}



\end{document}